\documentclass[a4paper,fleqn,usenatbib]{mnras}

\usepackage{mathptmx}
\usepackage[figuresright]{rotating}
\usepackage[graphicx]{realboxes}
\usepackage{adjustbox}
\usepackage{epstopdf}

\usepackage[T1]{fontenc}
\usepackage{ae,aecompl}

\usepackage{graphicx}	
\usepackage{amsmath}	
\usepackage{amssymb}	
\usepackage{pdflscape}

\title[The X-ray emission in young radio AGNs]{The X-ray emission in young radio AGNs}

\author[Mai Liao et al.]{Mai Liao,$^{1,2}$\thanks{E-mail: liaomai@shao.ac.cn}
Minfeng Gu,$^{1}$\thanks{E-mail: gumf@shao.ac.cn}
Minhua Zhou,$^{1,2}$\thanks{E-mail: zhoumh@shao.ac.cn}
Liang Chen$^{1}$\thanks{E-mail: chenliang@shao.ac.cn}
\\
$^{1}$Key Laboratory for Research in Galaxies and Cosmology, Shanghai Astronomical Observatory, Chinese Academy of Sciences,\\ 80 Nandan Road, Shanghai 200030, China \\
$^{2}$University of Chinese Academy of Sciences, 19A Yuquanlu, Beijing 100049, China \\
}

\date{Accepted XXX. Received YYY; in original form ZZZ}

\pubyear{2020}

\begin{document}
\label{firstpage}
\pagerange{\pageref{firstpage}--\pageref{lastpage}}
\maketitle

\begin{abstract}
In this work, we investigated the X-ray emission for a sample of young radio AGNs by combining their data from Chandra/XMM-Newton and at other wavebands. We find strong correlations between the X-ray luminosity $L_{\rm X}$ in 2$-$10 keV and the radio luminosities $L_{\rm R}$ at 5 GHz for VLBI radio-core, VLA radio-core and FIRST component, indicating that both pc- and kpc-scale radio emission strongly correlate with X-ray emission in these sources. We find approximately linear dependence of radio on X-ray luminosity in the sources with radiative efficient accretion flows (i.e., the Eddington ratio $R_{\rm edd} \gtrsim 10^{-3} $) with b $\sim$ 1 ($L_{\rm R}$ $\propto$ $L_{\rm X} ^{b}$) and  $\xi_{\rm RX}$ $\sim$ 1 in fundamental plane using VLBI data, where the dependence is consistent with the re-analysed result on the previous study in \cite{2016ApJ...818..185F} at $R_{\rm edd} \gtrsim 10^{-3}$, however is significantly deviated from the theoretical prediction of accretion flow as the origin of X-ray emission. In contrast to radio-quiet quasars, there is no significant correlation between $\Gamma$ and Eddington ratio. Our results seem to indicate that the X-ray emission of high-accreting young radio AGNs may be from jet. We constructed the SEDs for 18 sources (most are in radiative efficient accretion) including 9 galaxies and 9 quasars with high-quality X-ray data, and find that the X-ray emission of most quasars is more luminous than that of normal radio-quiet quasars. This is clearly seen from the quasar composite SED, of which the X-ray emission is apparently higher than that of radio-quiet quasars, likely supporting the jet-related X-ray emission in young radio AGNs. The scenario that the X-ray emission is from self-synchrotron Compton (SSC) is discussed.
\end{abstract}


\section{Introduction}

Young radio active galactic nuclei (AGNs) are characterized by compact radio structures and powerful radio emission (usually radio loud with radio loudness $R=L_{\rm 5GHz}/L_{\rm 4400 \AA}>10$,\citep{1989AJ.....98.1195K}) with characteristic convex radio spectrum. Usually, they can be divided into various subclasses according to convex frequency: high frequency peakers (HFP) radio sources that peak above 5 GHz (linear size (LS) $\leq$ 1 kpc), gigahertz peaked spectrum (GPS) radio sources with peak around 1 GHz (LS $\leq$ 1 kpc) and compact steep spectrum (CSS) radio sources peaking around 100 MHz (LS $\leq$ 20 kpc) \citep{od98}. Like normal AGNs, young radio AGNs are identified as quasars or galaxies. Galaxies show typical double symmetric or triple radio structures, while quasars display core-jet or complex morphologies. Quasars are frequently found at higher redshift compared to galaxies \citep{od98,paper1}, and they tend to own higher Eddington ratio than galaxies \citep{paper1}.
The population of these radio AGNs is thought to represent the early evolutionary stage of large-scale Fanaroff-Riley (FR) I/II \citep{1974FR} radio galaxies \citep{od98,2003PASA...20...38S}. This youth scenario is strongly supported by the measurements of dynamical and/or spectral age, about $10^{2}-10^{5}$ years \citep{1998A&A...336L..37O,2003PASA...20...19M,2003PASA...20...69P,2009AN....330..193G,2012ApJS..198....5A}.

While the previous studies on young radio AGNs mostly focused on radio band, the investigations on X-ray emission have been gradually increasing in recent years \citep[e.g.,][]{2006MNRAS.367..928V,2008ApJ...684..811S,teng09,2014MNRAS.437.3063K,2016ApJ...823...57S,2017ApJ...851...87O,2019arXiv190902084S}. However, the mechanism of X-ray emission is still in debate. In previous works, the X-ray emission has been proposed to be thermal comptonization emission from disc-corona system based on the observational perspective by systematically analyzing the spectral index, obscuration and comparison with larger AGNs\citep[e.g.,][]{2008ApJ...684..811S,teng09}. On the contrary, the spectral energy distribution (SED) modeling shows that the X-ray emission can be likely produced from non-thermal inverse Compton scatter process in either radio lobes for galaxies \citep{2008ApJ...680..911S,2010ApJ...715.1071O} or jets for quasars \citep{2004MNRAS.347..632W,2012ApJ...749..107M,2014ApJ...780..165M}. Moreover, the thermal emission observed from ISM caused by the interaction between jet and the surrounding materials has been reported in some sources  \cite[e.g., 3C 303.1 and 3C 305,][]{2017ApJ...851...87O}.

It has been proved that the relationship between the radio and X-ray luminosity ($L_{\rm R}$ $\propto$ $L_{\rm X} ^{b}$), and the fundamental plane of black hole activity (log $ L_{\rm R} = \xi_{\rm RX} $ log $ L_{\rm X} + \xi_{\rm RM} $ log $M_{\rm BH} + \rm constant$) are useful approaches to study the mechanism of X-ray emission in AGNs \cite[e.g.,][]{2003MNRAS.345.1057M}. Previous studies showed that $L_{\rm 5 GHz} \propto L^{\sim (0.6 - 0.7)}_{2-10 \rm keV}$ in low-luminosity AGNs (LLAGNs) and at underluminous low/hard spectra state of black hole X-ray binaries (BHBs) \citep[e.g.,][]{2003MNRAS.345.1057M,2006A&A...456..439K}, while the correlation becomes steeper $L_{\rm 5 GHz} \propto L^{\sim 1.4}_{\rm 2-10 keV}$ in luminous AGNs and at luminous low/hard spectra state of BHBs \citep{2011MNRAS.414..677C,2014ApJ...787L..20D}. The different radio$ - $X-ray slopes could be explained by the different accretion modes, where slopes of 0.7 and 1.4 correspond to radiatively inefficient and efficient accretion, respectively. The slope of radio/X-ray luminosity correlation in radio-loud AGNs appears to be much steeper compared with that of radio-quiet AGNs \citep{2008ApJ...688..826L,2011MNRAS.415.2910D}, possibly due to the jet contribution at X-ray band.

\cite{2016ApJ...818..185F} presented the first systematic study of radio/X-ray relation and fundamental plane for young radio AGNs consisting of high-excitation and low-excitation galaxies. The used radio data in their work was from NRAO VLA Sky Survey (NVSS, at $45 \arcsec$ resolution), while the X-ray data and black hole mass were collected from the literature. Their results on the radio/X-ray connection and fundamental plane are consistent with the predication of \cite{2003MNRAS.345.1057M}, implying that the X-ray radiation comes from the disk-corona system, rather than the jet. In this work, we aim to re-investigate the radio/X-ray relation and fundamental plane by constructing a larger sample, and making use of higher-resolution radio data (VLBI, VLA, and/or FIRST), to explore the origin of X-ray emission in young radio AGNs. We also compile multi-band data to study their global radiation properties. Section 2 shows the sample selection and data used in the work. In Section 3, our results are presented. Section 4 gives discussions and the main results are summarized in Section 5. Throughout the paper. the cosmological parameters $H_0 = 70\, \mathrm{km\,s^{-1}\, Mpc^{-1}}$, $\Omega_\mathrm{m} = 0.3$, and $\Omega_{\lambda} = 0.7$ are adopted. The spectral index $\alpha_{\nu}$ is defined as $f_{\nu}$ $\propto$ $\nu^{-{\alpha}_{\nu}}$ with $f_{\nu}$ being the flux density at frequency $\nu$.

\section{Sample and data}

\subsection{Sample selection}

To build the largest sample with X-ray data, we started with the parent sample of 468 young radio AGNs in \cite{paper1}, which was collected from available radio-selected samples in the literature and the blazar-type objects have been excluded. The sample was cross-matched with Chandra and XMM-Newton X-ray archives to search the available X-ray observations within 2 $\arcsec$ and 5 $\arcsec$ to NASA/IPAC Extragalactic Database (NED) source positions, respectively. We found that 89 out of 468 sources have available X-ray data from Chandra and/or XMM-Newton observations. Additional two CSS radio sources of 3C 305 \citep{2009ApJ...692L.123M} and 4C 13.66 \citep{2013ApJ...773...15W} were also included. Our final X-ray sample thus consists of 91 sources, which are listed in Table 1. It should be noticed that our sources are heterogeneous and their properties are possibly strongly influenced by different selection purpose in the literature. Nevertheless, it's the largest sample with X-ray observations, enabling us to study the emission in young radio AGNs.

\subsection{Data}
\subsubsection{X-ray}
The X-ray data from Chandra and XMM-Newton were searched from the literature. When both Chandra and XMM-Newton data are available for the source, in principle the Chandra data is preferred due to its higher spatial resolution. But for 2MASX J19455354+7055488, PKS2127+04, COINS J2022+6136, we used XMM-Newton data because of their deeper observations and better spectral analyses than those of Chandra in the literature. Six sources were excluded in the analysis due to various reasons. Two of them are lensed sources, CGSRaBs J1424+2256 \citep{1992MNRAS.259P...1P} and 4C+05.19 \citep{Schechter1993}. It's difficult to determine their intrinsic emission at X-ray and radio bands. NGC 262 was excluded because the significant X-ray spectral variability was reported in the literature \citep{2015A&A...579A..90H}. The X-ray emission of 4C 31.04 and SDSS J124733.31+672316.4 \citep{2016ApJ...823...57S}, and SDSS J130941.51+404757.2 are too faint to be measured from Chandra or XMM-Newton observations. After excluding these six objects, the X-ray data is available in 85 sources.

In order to study the X-ray emission in details, both the flux density at 2 $-$ 10 keV and photon index $\Gamma$ are required from the literature. 
For our sample of 85 sources, these information are available for 56 objects, of which the luminosity at 2 $-$ 10 keV can be calculated. 
In four sources (4C+00.02, PKS 0428+20, 4C+14.41 and PKS 2008-068), the detailed X-ray spectral analysis can't be performed due to low signal-to-noise ratio \citep{teng09}. Therefore, only the 2 $-$ 10 keV  luminosities were directly taken from \cite{teng09}, which was either calculated with assumed photon index (PKS 0428+20, 4C+14.41 and PKS 2008-068), or extrapolated from the soft X-ray band (4C+00.02).

The X-ray data of the remaining 25 sources (twenty-two 3C sources, PKS B1421-490, PKS 1413+135, and PKS 0237-23) have been analyzed and published in various papers \citep{2010ApJ...714..589M,2011ApJS..197...24M,2012ApJS..203...31M,2012MNRAS.424.1774H,2013ApJS..206....7M,2015ApJS..220....5M,2018ApJS..235...32S,2010ApJ...723..935H}.
However, these works mainly focused on either only the flux \cite[e.g.,][]{2015ApJS..220....5M} or the extended X-ray emission. 
In order to get both the flux and photon index, we re-analyzed their Chandra and XMM-Newton data.
We used Chandra Interactive Analysis of Observations software (CIAO) v4.8 and Chandra Calibration Database (CALDB) version 4.7.1 to reduce
the data \citep{2006SPIE.6270E..1VF}. $Chandra\_repro$ script was used to create a new level 2 event file and a new bad pixel file. Data energy was filtered between $ 0.3-10\rm ~keV $, and the background flares were carefully checked. We extracted the spectrum of core region within a radius of 2.5$\arcsec$ with $ specextract $ script, and a $20-30\arcsec$ source-centered annulus was used for background region. We carefully calculated pile-up fraction with PIMMS for all Chandra data, and found no data affected  by pile-up effect.

PKS $ 0237-23 $ has only XMM-Newton data (observation ID: 0300630301). We processed the data with Scientific Analysis Software (SAS) package step by step with SAS cookbook\footnote{\url{https://heasarc.gsfc.nasa.gov/docs/xmm/abc/}}. The X-ray spectrum was extracted from a source-centered radius of $32\arcsec$ with background region $40\arcsec$ near the source, and it was later binned by minimum 25 counts for background-subtracted spectral channel. We also checked pile-up effect with $epatplot$, and no pile-up effect was found. 

All the extracted spectra were fitted with Xspec using an absorbed power-law model with absorption components of galactic absorption \citep{2005A&A...440..775K} and intrinsic absorption, which are characterized by the equivalent neutral hydrogen column density of $N^{\rm gal}$ and $N_{\rm H}^{z_{\rm abs}}$. The detailed spectral fit can be carried out in 19 objects with counts larger than 40 within a radius of 2.5$\arcsec$, then the luminosity and photon index can be obtained. 
In the remaining six objects with low counts, a fixed photon index of $\Gamma=1.7$ was used to calculate the 2 $-$ 10 keV X-ray luminosity. The X-ray luminosity and photon index for our sample is shown in Table 1.  

\subsubsection{Radio}

We carefully searched the published VLBI and VLA data at 5 GHz only for radio core from NED and the literature, with the aim to minimize the contamination from extended emission of radio lobes or host galaxies. When 5 GHz VLBI observations are not available, we collected 8.4 GHz VLBI radio-core emission in five sources (see Table 1). Among 91 sources, the VLBI data were obtained for 44 sources, and the VLA data is available for 52 objects which are usually unresolved in VLA images showing compact structure and point-like morphology. As supplementary, the 1.4 GHz flux is also included in our work for 48 objects from Faint Images of the Radio Sky at Twenty-Centimeters (FIRST) \citep{2015ApJ...801...26H}. And also, our sources are unresolved in FIRST images. The K-correction was applied to calculate the luminosity assuming the spectra index of 0 at 5 GHz, and 0.5 at 1.4 GHz.

\subsubsection{Optical/UV}
The optical spectra of our sample were collected from SDSS DR12. The black hole mass and Eddington ratio were directly taken from \cite{paper1}, in which the spectroscopic analysis was carried out on the largest optical sample of young radio AGNs using the data from SDSS DR12. The BH masses were estimated with various empirical relations utilizing the line width and luminosity of broad lines (i.e., H$\beta$, Mg II, C IV) in order to avoid the contamination from non-thermal jet emission, or with the relation of BH mass and stellar velocity dispersions ($\sigma_{\ast}$), or with [OIII] line width which was used as surrogate of $\sigma_{\ast}$ (see details in \cite{paper1}). The Eddington ratio $R_{\rm edd}= L_{\rm bol}/ L_{\rm Edd}$ was calculated with the bolometric luminosity $L_{\rm bol}$ estimated from the luminosities of broad lines (i.e., $L_{\rm H \beta}$, $L_{\rm Mg II}$, $L_{\rm C IV}$) or narrow [OIII] line, and the Eddington luminosity $L_{\rm edd} = 1.38 \times 10^{38} M_{\rm BH}/\rm M_{\rm \sun}$ \cite[see details in][]{paper1}. 3C 305 is not included in the original sample of \cite{paper1}, and its SDSS spectrum was analyzed as in \cite{paper1}, from which the black hole mass and Eddington ratio were then calculated. 

In addition, we collected the FUV and NUV data for our sample from GALEX \citep{2007ApJS..173..682M}.

\subsubsection{Infrared}

To build spectral energy distribution for young radio AGNs, the infrared data was collected from near-infrared survey of Two Micron All Sky Survey \cite[2MASS,][]{2006AJ....131.1163S} at 1.2, 1.6, and 2.2 $\micron$, and from United Kingdom Infrared Telescope (UKIRT) Infrared Deep Sky Survey \cite[UKIDSS,][]{2012yCat.2314....0L} at 1.0, 1.2, 1.6, and 2.2 $\micron$. Moreover, the mid-infrared photometry data from Wide-field Infrared Survey Explorer \cite[WISE,][]{2010AJ....140.1868W} at 3.4, 4.6, 12, and 22 $\micron$, and from Infrared Array Camera (IRAC) at 3.6, 4.5, 5.8, and 8 $\micron$ \citep{2004ApJS..154...10F}, together with Multiband Imaging Photometer for Spitzer \cite[MIPS;][]{2004ApJS..154...25R} at 24, 70, and 160 $\micron$ were also collected when available.

\subsubsection{Sample properties}
\begin{figure*}
	\centering
	\includegraphics[width=3.4in]{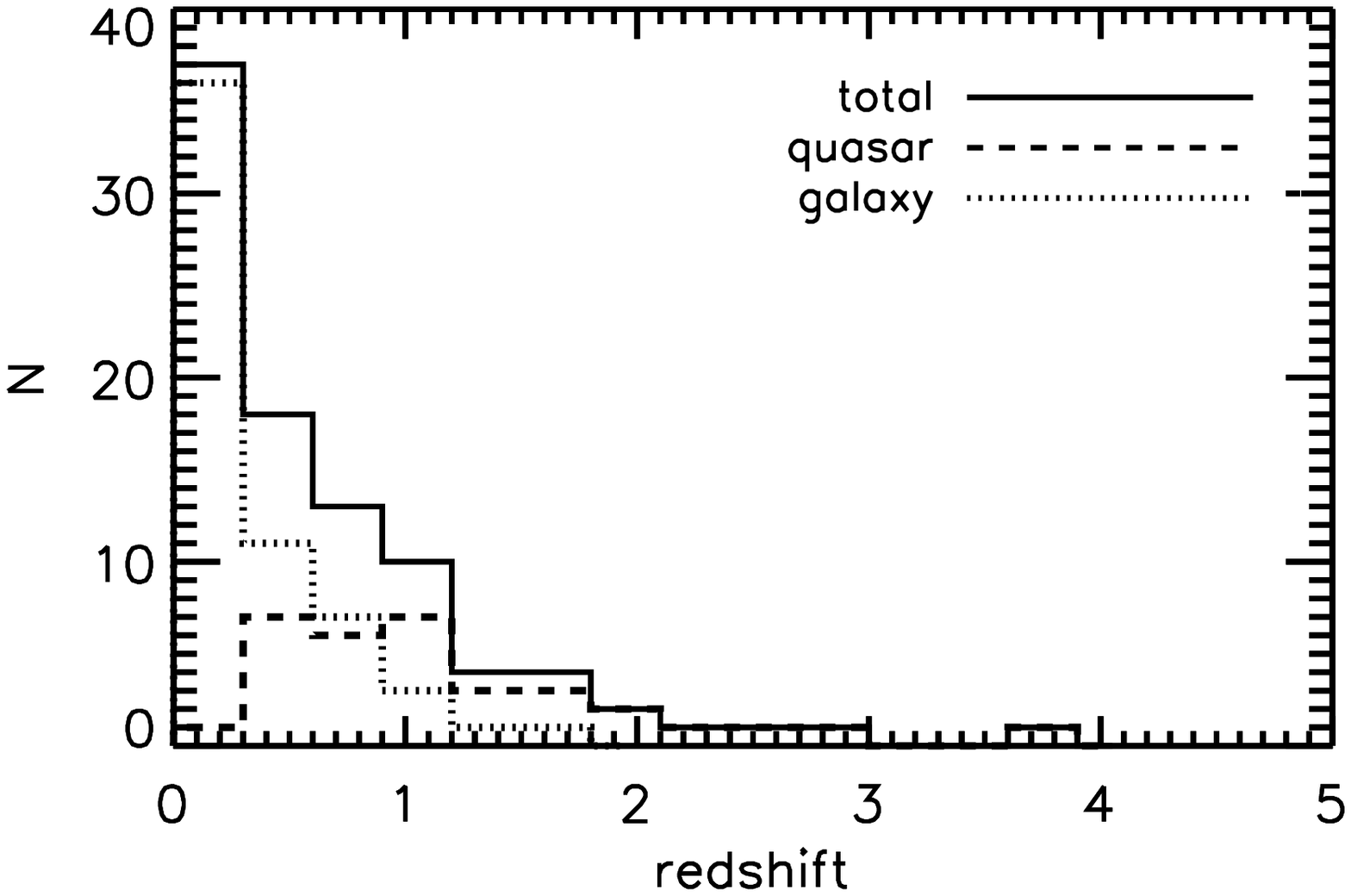}
	\includegraphics[width=3.4in]{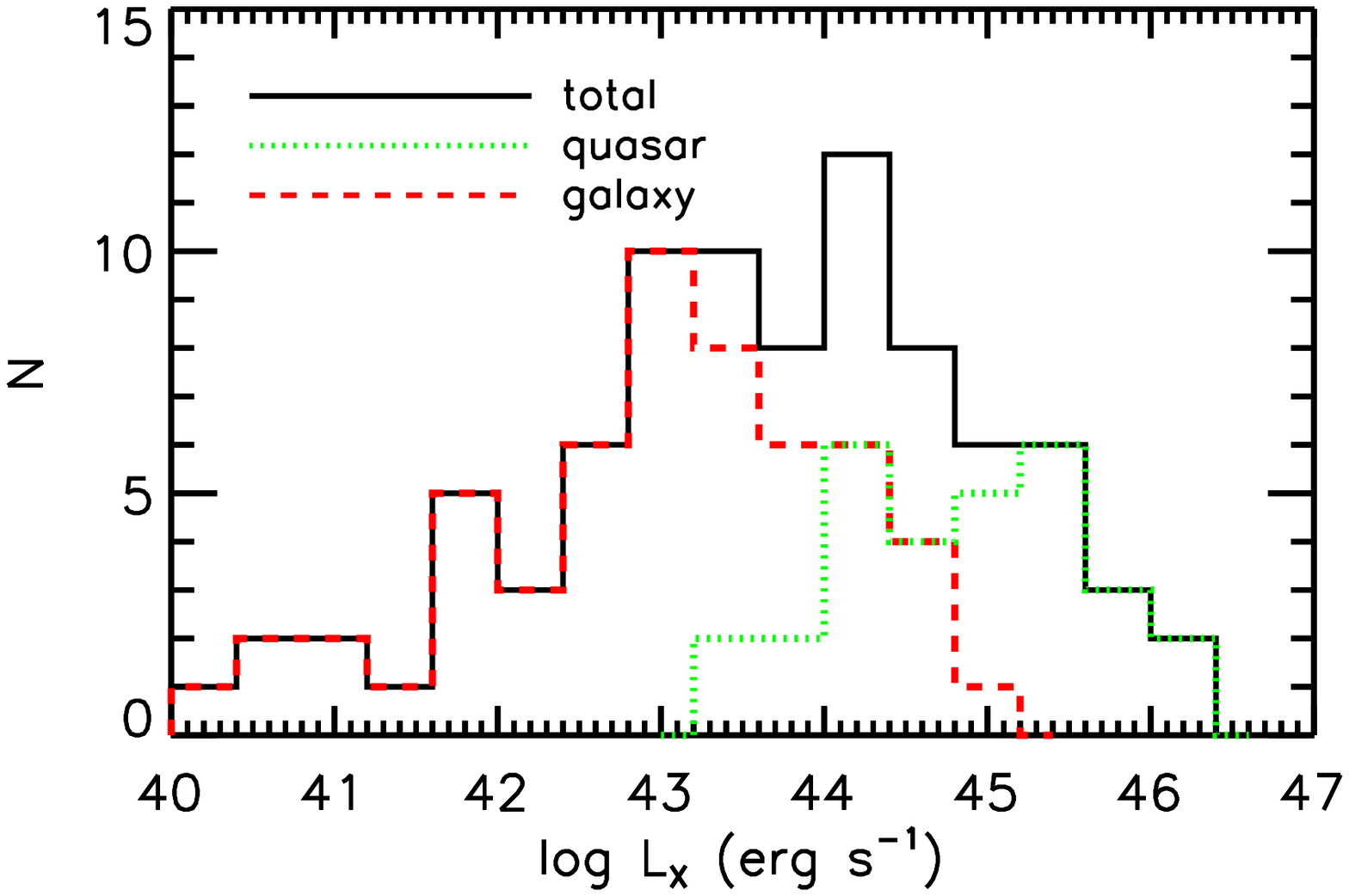}
	\includegraphics[width=3.4in]{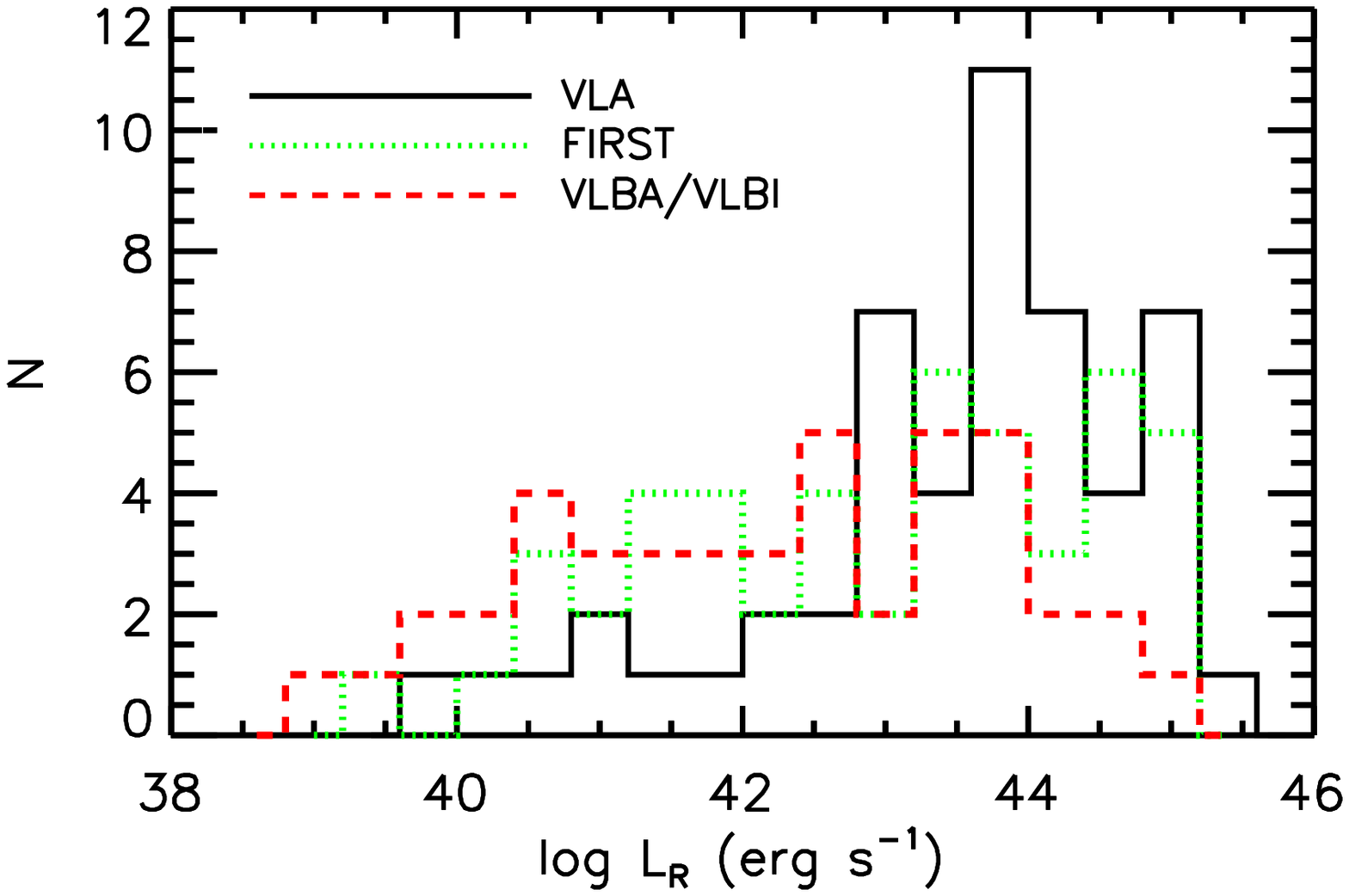}
	\includegraphics[width=3.4in]{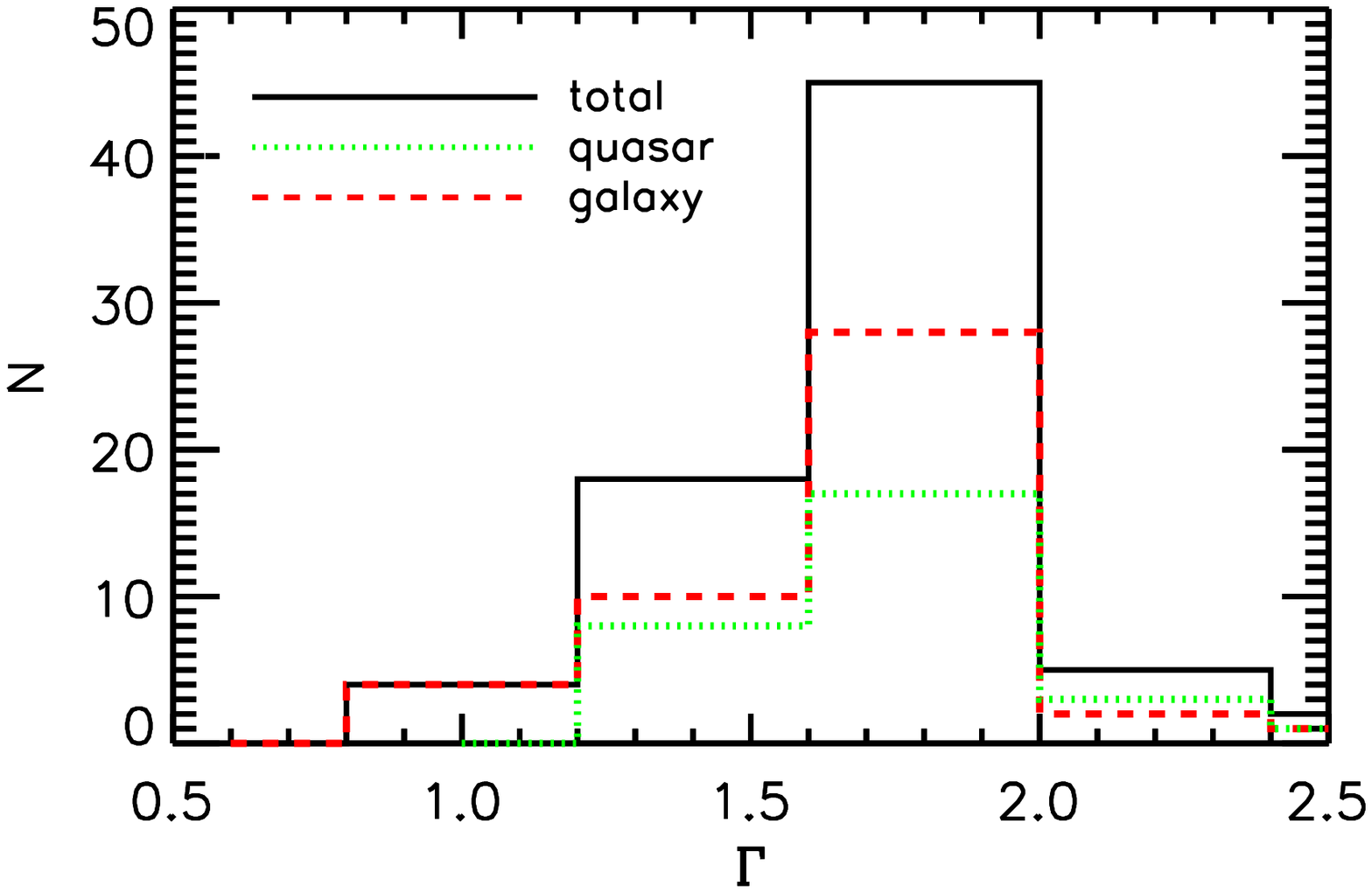}
	\vspace*{-0.2 cm} \caption{The distributions of various parameters for the sample: $upper~ left:$ redshift; $upper~ right:$ the X-ray luminosity at 2$-$10 keV; $lower~ left:$ the radio luminosity at 5 GHz; $lower~ right:$ the X-ray photon index in 2$-$10 keV. \label{sample}}
	\vskip-10pt
\end{figure*}

Our sample of 91 sources is shown in Table 1, consisting of 52 CSS, 26 GPS, 4 HFP and 9 compact symmetric objects (CSO). The sample sources were further classified as 57 galaxies and 34 quasars. The distributions of redshift, X-ray luminosity at 2 $-$10 keV, radio luminosity at 5 GHz, and X-ray photon index are shown in Figure \ref{sample}. The X-ray luminosities of quasars are systematically higher than those of galaxies, with galaxies covering about five order of magnitude at lower luminosity $10^{40}$ $-$ $10^{45}$ $\rm erg\ s^{-1}$ compared to quasars mainly at $10^{43}$ $-$ $10^{46}$ $\rm erg\ s^{-1}$.
The VLA/FIRST luminosity is systematically larger than VLBI luminosity by about one order of magnitude when the sources with both VLA/FIRST and VLBI data considered (i.e., $L_{\rm VLA/FIRST} ~\sim$ 10 $L_{\rm VLBI}$). This suggests that large scale (arcsec scale) emission is dominant in VLA and FIRST images, although they are usually compact. We find that the photon index $\Gamma$ at 2 - 10 keV has a broad distribution from 0.86 to 2.62 with median and mean values $\sim1.7$. There is no significant difference in $\Gamma$ distribution between galaxies and quasars. It should be noted that only free-fitting derived $\Gamma$ in X-ray spectra analysis either in the literature or from our data analysis was contained in our following statistic analysis.

\section{Results}
\subsection{Radio/X-ray correlation}

\begin{figure*}
	\centering
	\includegraphics[width=6.0in]{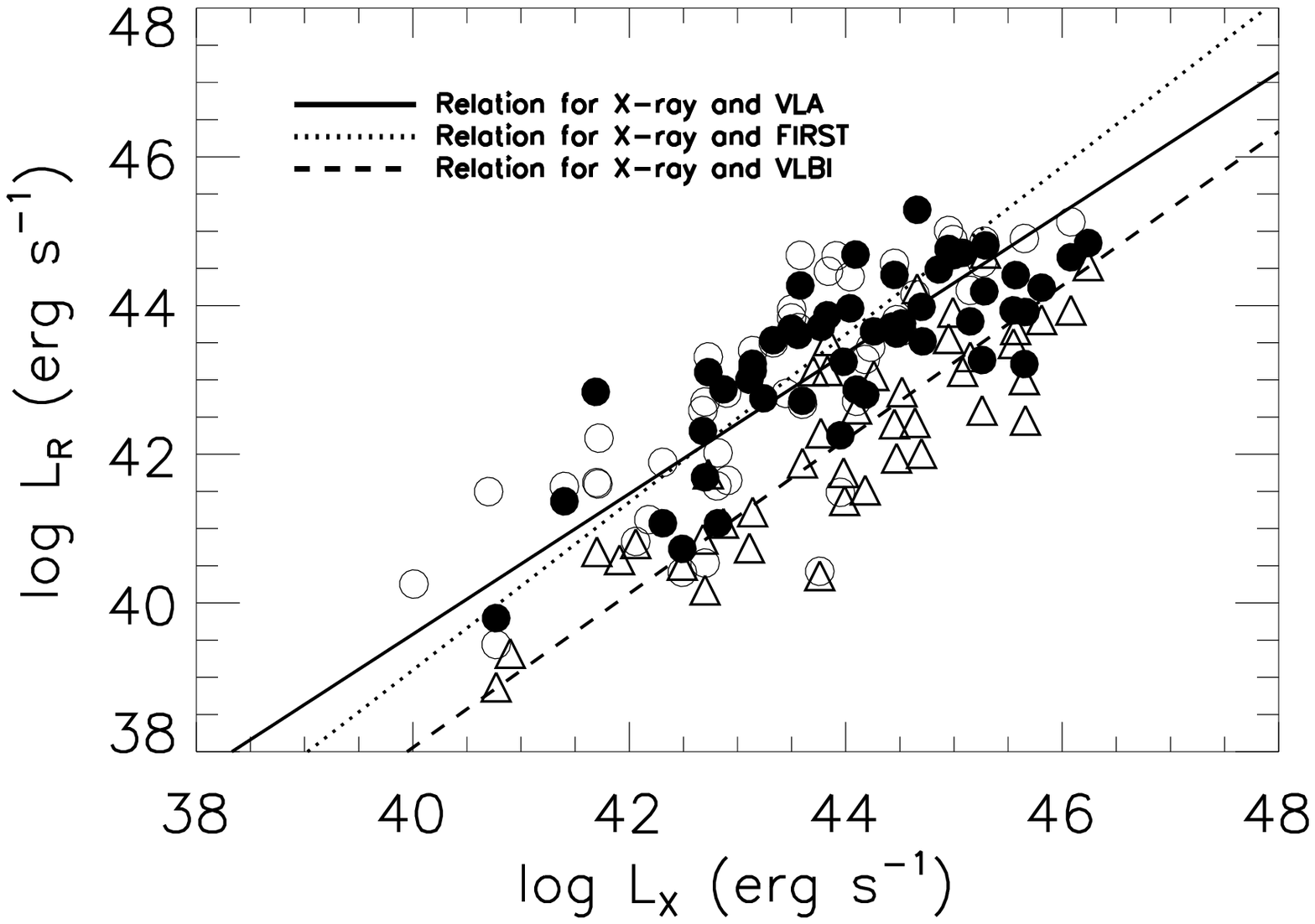}
	\vspace*{-0.2 cm}\caption{The radio luminosity at 5 GHz and the X-ray luminosity at 2-10 keV. The solid, dotted and dashed lines are linear fits using OLS bisector method for VLA, FIRST and VLBI data, respectively.\label{Lx_Lr}}
	\vskip-10pt
\end{figure*}

In Figure \ref{Lx_Lr}, the X-ray luminosity is plotted with the radio luminosity for 47 FIRST sources, 50 VLA core, and 42 VLBI core detections. We found strong correlations between the X-ray luminosity $L_{\rm X}$ at 2$-$10 keV and the radio luminosity $L_{\rm R}$ at 5 GHz of VLA, FIRST, and VLBI, with Spearman rank correlation coefficient $r_{\rm s}=$ 0.77, 0.82, 0.84, respectively, and all with the probability $P_{\rm null} < 10^{-10}$ for the null hypothesis of no correlation. The partial Kendall $\tau$ correlation test \citep{1996MNRAS.278..919A} was performed to exclude the common dependence of luminosities on the redshift. The partial correlation between $L_{\rm X}$ and $L_{\rm R}$ is still significant ($P_{\rm null} < 10^{-3}$). The strong correlations imply that both pc- and kpc-scale radio emission have tight relation with X-ray emission in our sample.
The Ordinary Least Squares bisector (OLS bisector) \citep{1990ApJ...364..104I} fits to the correlations between $L_{\rm X}$ and $L_{\rm R}$ give:

\begin{equation}
\log L_{\rm R_{\rm VLA}} = 0.95(\pm0.08) \times \log L_{\rm X}+1.79(\pm3.62)
\end{equation}

\begin{equation}
\log L_{\rm R_{\rm FIRST}} = 1.13(\pm0.09) \times \log L_{\rm X}-6.04(\pm3.97)
\end{equation}

\begin{equation}
\log L_{\rm R_{\rm VLBI}} = 1.04(\pm0.06) \times \log L_{\rm X}-3.37(\pm2.83)
\end{equation}
All these fits show approximately linear relations between log $L_{\rm X}$ and log $L_{\rm R}$. 

\subsection{Fundamental plane}

The fundamental plane can be used to constrain the mechanism of X-ray emission. The dependence of radio on X-ray emission can be quite different between radiative efficient and inefficient accretion flows. In our sample, 34 objects have both BH mass and Eddington ratio estimations. Using a dividing value of Eddington ratio 0.001 to distinguish radiative efficient and inefficient accretion flows, the radiative efficient flow ($R_{\rm edd} \ga 10^{-3} $) \citep{2015MNRAS.448.1099Q} is present in 26 sources. It provides us an opportunity to explore the fundamental plane of young radio AGNs under the condition of radiatively efficient accretion. The source number reduces to 13 when VLBI radio-core data is required, where using VLBI data can effectively exclude the dominant large-scale radio emission in FIRST/VLA data as shown in Section 2.2.2 and 2.2.5. We utilized the multiple linear regression of Bayesian approach \citep{2007ApJ...665.1489K} to investigate the fundamental plane as used in \cite{2016ApJ...818..185F}, i.e. the relation of $L_{\rm R}$, $L_{\rm X}$, and $M_{\rm BH}$. We adopted the typical uncertainties $\sigma_{R}$ = 0.2, and $\sigma_{X}$ = 0.3 \cite[e.g.,][]{2018MNRAS.481L..45L}. The uncertainty of black hole mass $\sigma_{M}$ was estimated from the measurement uncertainties of various line parameters in empirical relation in \cite{paper1}. 
We performed 5000 times Monte Carlo simulation with uncertainties in radio and X-ray luminosity varying between 0.02 and 0.2 dex, and 0.02 and 0.3 dex, respectively. 
The median values of 5000 times experiments were taken as the fitted results for each parameters, and the intervals containing 68\% were considered as the corresponding errors.

Our best fitting results show:
\begin{equation}
\log L_{\rm R_{\rm VLBI}}=0.99^{+0.05}_{-0.05} \log L_{\rm X}-0.15^{+0.06}_{-0.06}  \log M_{\rm BH} + 0.04^{+1.70}_{-1.70} 
\end{equation}
with intrinsic scatters $\sigma$ = 0.57.\\
\\
The corresponding linear relation (OLS bisector) between $L_{\rm X}$ and $L_{\rm R}$ are:
\begin{equation}
\log L_{\rm R_{\rm VLBI}}=1.02(\pm 0.14) \log L_{\rm X}-2.98(\pm 6.07)
\end{equation}

As comparison, we also study the fundamental plane when large scale radio emission included by using FIRST data as did for VLBI data \footnote{The FIRST data can better represent kpc radio emission than VLA data due to the consistent resolution.}. There has 12 sources with both VLBI radio-core and FIRST detections, and $R_{\rm edd} \ga 10^{-3} $. The best fitting results show:

\begin{equation}
\log L_{\rm R_{\rm FIRST}}=1.08^{+0.07}_{-0.07} \log L_{\rm X}+0.11^{+0.14}_{-0.14} \log M_{\rm BH} -5.10 ^{+1.84}_{-1.84} 
\end{equation}
with intrinsic scatters $\sigma$ = 0.67.

\begin{equation}
\log L_{\rm R_{\rm FIRST}}=1.15(\pm 0.23) \log L_{\rm X}-7.44(\pm 10.52)
\end{equation}

Our results show that b $\sim$ 1 ($L_{\rm R}$ $\propto$ $L_{\rm X} ^{b}$) and  $\xi_{\rm RX}$ $\sim$ 1 for the high accreted sources in our sample with pc-scale radio emission. While $\xi_{\rm RX}$ in FIRST data is slightly steeper than that in VLBI data, the values of $\xi_{\rm RX}$ are consistent with each other within errors. 

The fundamental plane and radio/X-ray relation for VLBI radio core and FIRST radio luminosity in our high-accreted objects are shown in Figures \ref{vlba} and \ref{first}. When available, the uncertainty on $L_{\rm X}$ is in the range of 0.04 $-$ 0.29 dex, and the uncertainty of VLBI $L_{\rm R}$ is 0.03 $-$ 0.11 dex. In constrast, the uncertainty of FIRST $L_{\rm R}$ is rather small ($< 10^{-3}$ dex), and can be usually neglected.

\begin{figure*}
	\centering
	\includegraphics[width=3.4in]{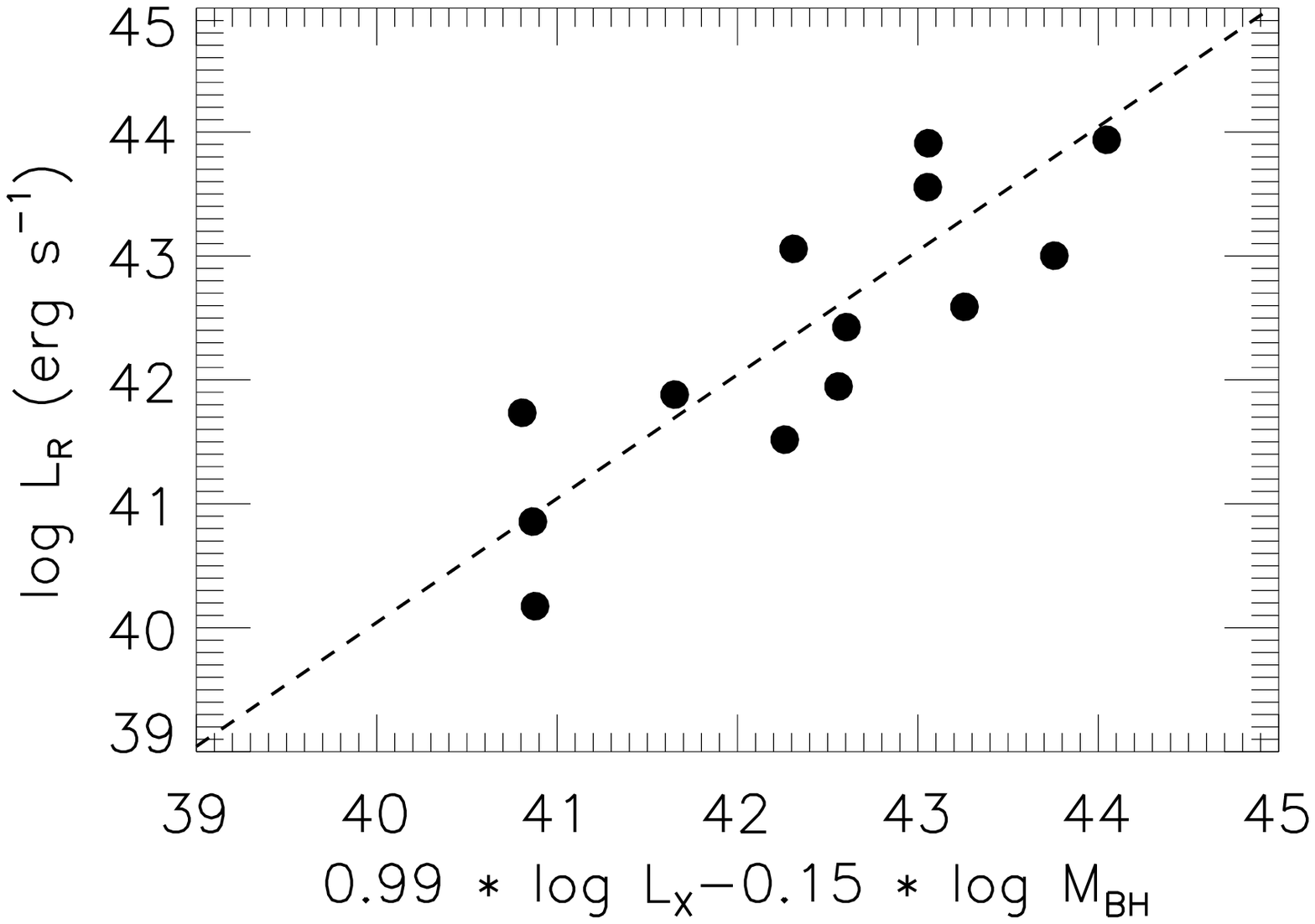}
	\includegraphics[width=3.4in]{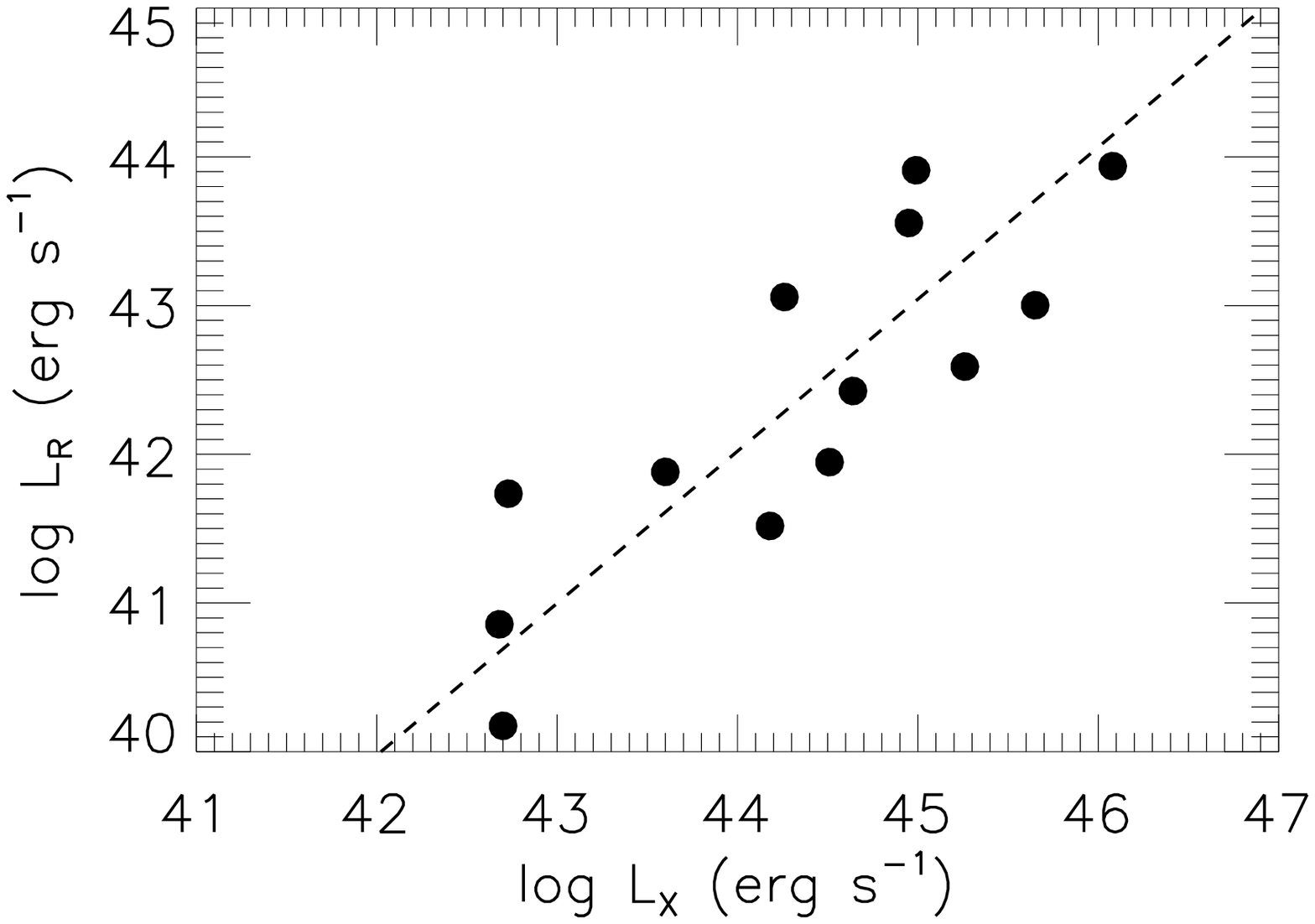}
	
	\vspace*{-0.2 cm} \caption{The fundamental plane of black hole activity for young radio AGNs with VLBI radio-core detection and $R_{\rm edd} \ga 10^{-3} $ ($left$). The dashed line is the best fit; The relationship between $L_{\rm R}$ and $L_{\rm X}$ in young radio AGNs with VLBI radio-core detection and $R_{\rm edd} \ga 10^{-3} $ ($right$). The dashed line is the best fit using OLS bisector method. \label{vlba}}
	\vskip-10pt
\end{figure*}\label{vlba}

\begin{figure*}
	\centering
	\includegraphics[width=3.4in]{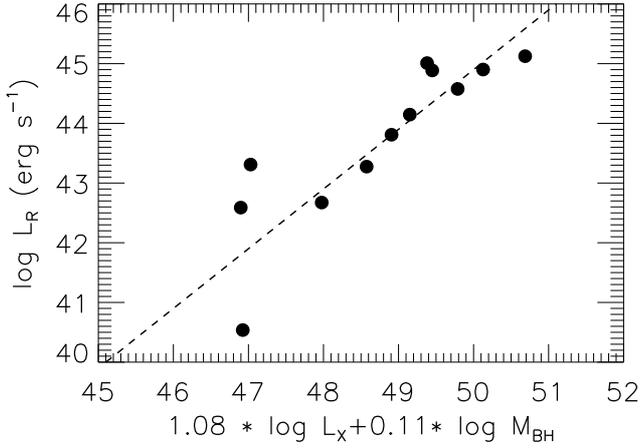}
	\includegraphics[width=3.4in]{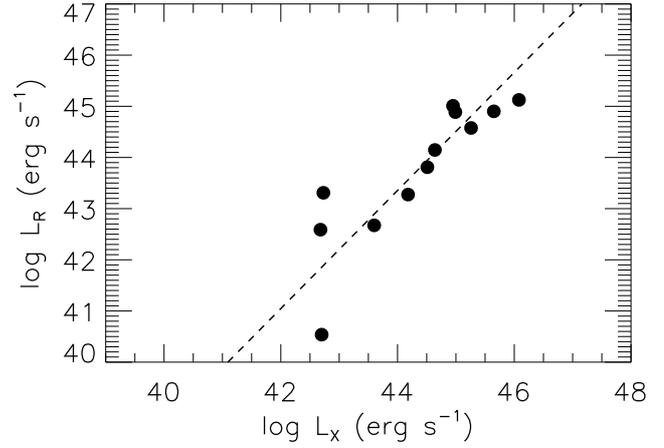}
	
	\vspace*{-0.2 cm} \caption{Same as Figure \ref{vlba}, but for the sources with FIRST detections. \label{first}}
	\vskip-10pt
\end{figure*}\label{first}

\subsubsection{$\Gamma$ vs Eddington ratio}
\begin{figure}
	\centering
	\includegraphics[width=3.4in]{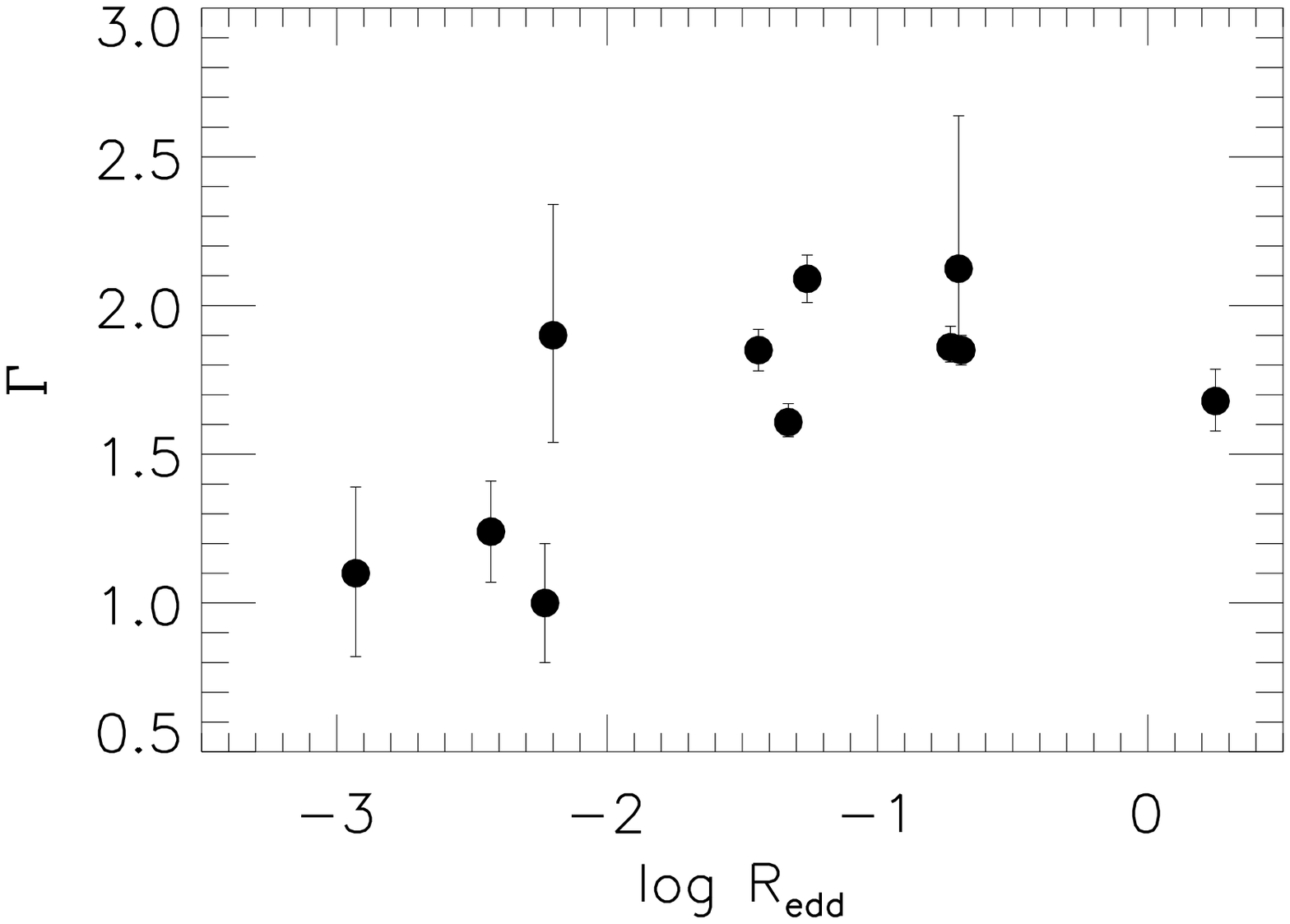}
	
	\vspace*{-0.2 cm} \caption{The photon index $\Gamma$ versus the Eddington ratio for sources with $R_{\rm edd} \ga 10^{-3} $ and radio detections with both VLBI core detections and FIRST detections. \label{gamma_logR}}
	\vskip-10pt
\end{figure}

The significant correlation between the hard X-ray photon index $\Gamma$ and Eddington ratio has been found in radio-quiet AGNs in the literature \citep{2015A&ARv..23....1B,2019MNRAS.490.3793L}, which would be the clue on the X-ray emission from disk-corona system.
\cite{2019MNRAS.490.3793L} found that this correlation is very strong in radio-quiet AGNs, while no significant correlation is present in radio-loud AGNs. The authors argued that the X-ray emission from the jet in radio-loud AGNs should be very important. Thus, studying the relationship between $\Gamma$ and Eddignton ratio $R_{\rm edd}$ may shed light on the X-ray emission for our high-accreted sources, which is shown for sources with VLBI core detections and FIRST measurements in Figure \ref{gamma_logR}, where
the uncertainty on $\Gamma$ is less than 0.20 in most sources (see Table 1). We failed to find any significant correlations ($P_{\rm null} > 0.05$), as \cite{2019MNRAS.490.3793L}. 

\subsection{Spectral energy distribution}
Besides the correlation analysis on multi-band emission, the spectral energy distribution (SED) can also be used to study the nature of multi-band emission for young radio AGNs, especially when compared with other AGN populations. The quasar composite SED \cite[e.g.,][hereafter S11]{2011ApJS..196....2S} shows that the prominent differences between radio-loud quasar (RLQ) and radio-quiet (RQQ) quasar are in radio and X-ray bands but have almost same feature in other bands . The distinction in radio band could be most likely due to the presence of jets in RLQs. The reason of difference at X-ray band is unclear, which may be caused by the powerful jets in RLQs. Comparing the composite SED of young radio AGNs with normal quasars may shed light on their global SEDs, especially at X-ray band.

In order to well study the X-ray spectra, we only considered the sources with X-ray spectral fitting at high significance either in the literature or from our data analysis.
In this case, 56 sources in our sample are included, of which 27 objects have available optical spectrum in SDSS DR12.
It has been shown that the optical/UV composite spectrum of young CSS quasars may suffer from the dust reddening and show steep continuum and large Balmer decrements \citep{1995baker}. While the dust extinction is likely present at optical/UV bands, the infrared (IR) band is less affected, which can be used as normalization in constructing the composite SED.  
We searched the IR data covering the rest-frame $1 \micron$, and found 22 out of 27 sources have near-IR data from 2MASS/UKIDSS. 
The final source number reduces to 18 (9 quasars and 9 galaxies), after excluding four sources with weak nuclei emission in Chandra images (SDSS J092405.30+141021.4, SDSS J132419.67+041907.0, and 3C 305) or uncertain complicated X-ray spectrum (Mrk 0668).

\subsubsection{Individual objects}

\begin{figure*}
	\centering
	\includegraphics[width=6.0in]{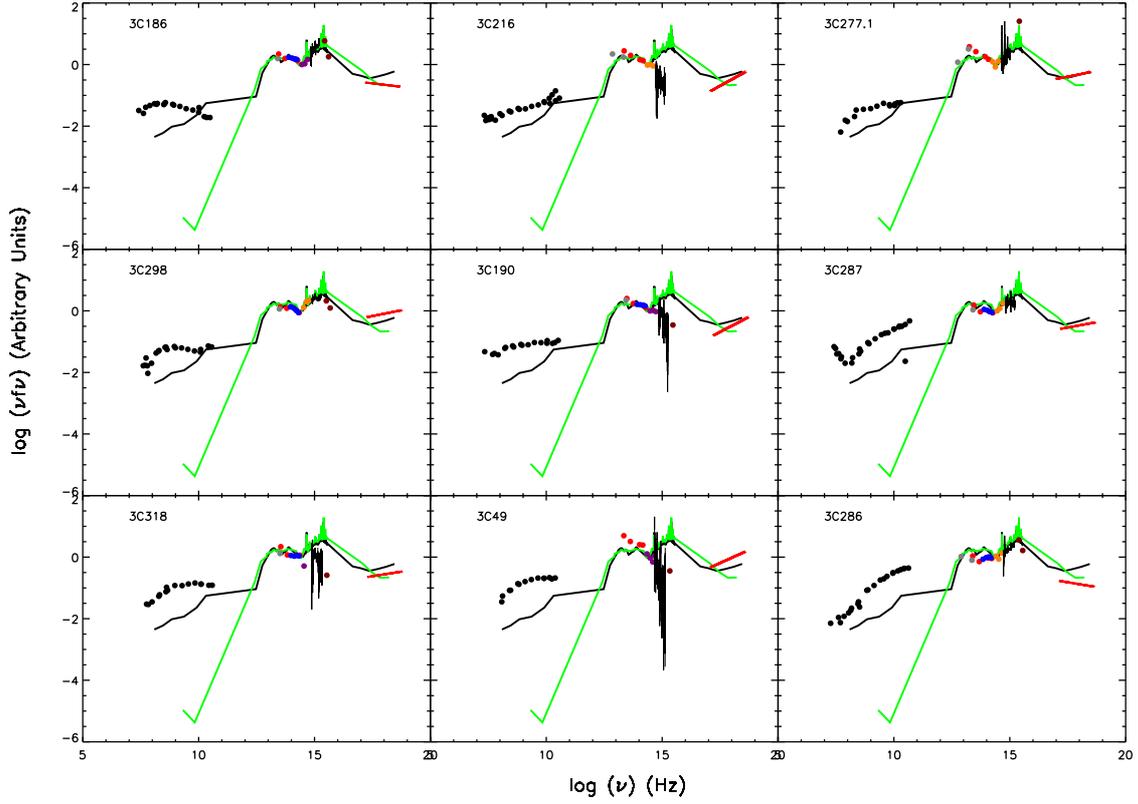}
	\vspace*{-0.2 cm}\caption{The SEDs of nine quasars from radio to X-ray band. The black and green thick solid lines are composite SEDs of radio-loud and radio-quiet quasars in S11, respectively. All SEDs are normalized at 1 $\micron$. The circles stand for the multi-band data from radio to optical/UV. The red and black solid lines represent the X-ray and optical spectra, repectively.\label{qso}}
	\vskip-10pt
\end{figure*}

\begin{figure*}
	\centering
	\includegraphics[width=6.0in]{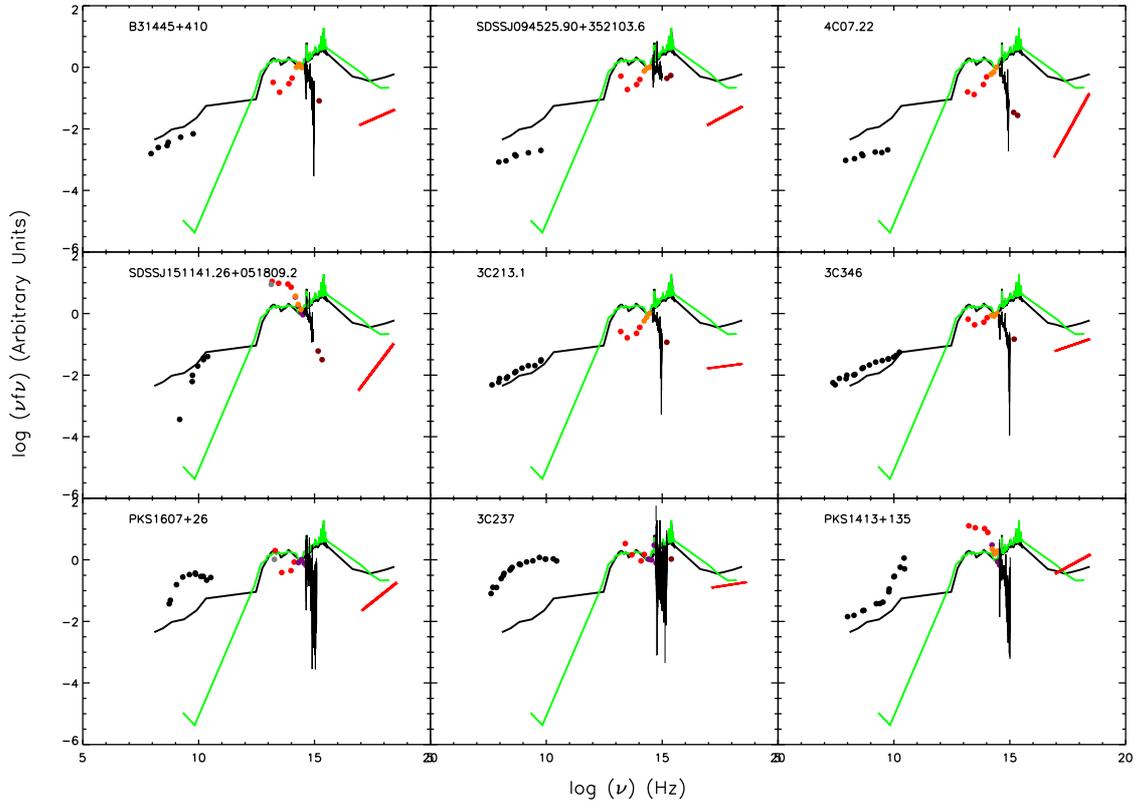}
	\vspace*{-0.2 cm}\caption{Same as Figure \ref{qso}, but for nine galaxies. \label{galaxy}}
	\vskip-10pt
\end{figure*}

The rest-frame broadband SED for individual objects in log $\nu f_{\nu} - \rm log ~\nu$ space from radio to X-ray bands are shown in Figures \ref{qso} and \ref{galaxy}, for quasars and galaxies, respectively, together with SEDs of RLQs and RQQs from S11 as comparison. The Eddington ratio in all quasars and most galaxies  are greater than $10^{-3}$. The multi-band flux density was normalized to rest-frame $1 \micron$ (log $\nu\sim 14.5$). The radio data from 10 MHz to 23 GHz collected from NED clearly shows prominent excess compared to RLQs in most sources, especially in all quasars. This is consistent with the notion that young radio AGNs are efficient radio emitters at early stage of radio activity \citep{paper1}, and the radio luminosity will decrease when they finally evolve to large-scale radio sources \citep{od98}. 
We find that the X-ray spectra at 0.3$-$10 keV of all quasars except for 3C 286 and 3C 186 are higher than that of S11 RQQs. In contrast, the X-ray emission of all galaxies except for PKS 1413+135 is below that of RQQs in S11. The infrared emission in quasars is in general consistent with that of S11, indicating a thermal IR bump mainly emitted from dusty torus. The IR bump is only seen in three galaxies with two even more significant (SDSS J151141.26+051809.2 and PKS 1413+135), and the rest one (3C 237) comparable to S11 quasars. As the indicator of thermal emission from accretion disc, the big blue bump (BBB) is visible in five quasars (3C 186, 3C 277.1, 3C 286, 3C 287 and 3C 298), however it's not seen in the rest of objects (3C 49, 3C 190, 3C 216 and 3C 318). As expected, BBB is lacking in all galaxies. Despite the SEDs of galaxies being different from those of quasars in optical, infrared and X-ray bands may due to the obscure caused by the larger view angle, the intrinsic emission for quasar and galaxy should be same under the AGN unification model.

\subsubsection{Composite SED}

The same method of constructing composite SED as in S11 was applied for our young radio quasars. We rebined the normalized data for each waveband, where the median value and the central frequency of each bin represent one flux point and one frequency of that point in the final composite SED, respectively. The use of median value in the composite SED can effectively prevent the extreme data point as suggested by S11. 
The adopted bin numbers for each band were determined by considering both the statistical significance and spectral features (e.g., S11). Eight bins were applied to rebin the radio data from 7.26 to 10.67 in log $\nu$ space with interval $\Delta~ \rm log~ \nu$ = 0.426. At infrared band, seven bins were used in log $\nu$ range of 12.75 $-$ 14.77 with $\Delta~ \rm log~ \nu =0.289$. The SDSS spectra were rebinned in 160 bins with a bin size of $\Delta~ \rm log~ \nu=0.0035$ in 14.63 $-$ 15.20, which is covered by most quasars. The NUV and FUV data from GALEX in the frequency range of 15.33 $-$ 15.67 were rebined in two bins with bin size of 0.17. At X-ray band, we defined 10 bins within 17.27 $-$ 18.50 frequency range, with a bin size of $\Delta~ \rm log~ \nu$ = 0.123. 

\begin{figure*}
	\centering
	\includegraphics[width=6.0in]{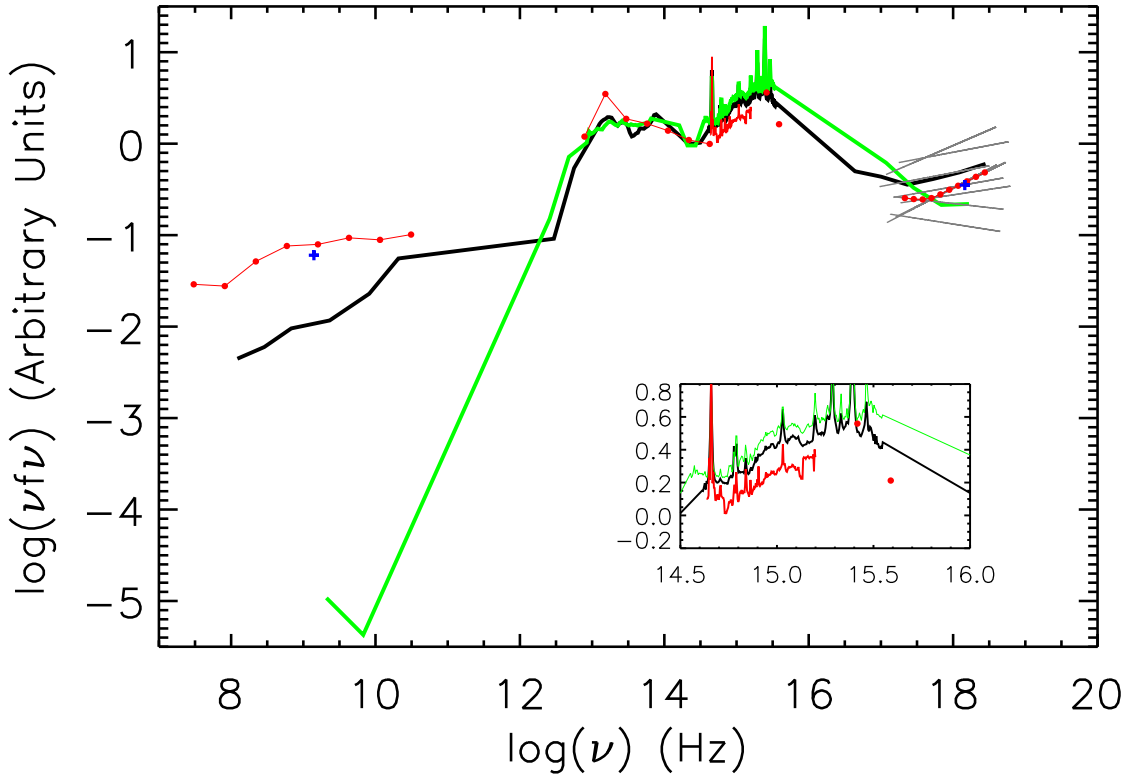}
	\vspace*{-0.2 cm}\caption{The median composite SED of our young radio quasars in red solid lines and circles. The black and green thick solid lines are composite SEDs of radio-loud and radio-quiet quasars in S11, respectively. All composite SEDs are normalized at 1 $\micron$. The grey lines are X-ray spectra for individual objects. The blue crosses are the median values at 1.4 GHz and 6 KeV for all quasars in our sample with near-infrared data covering the rest-frame 1$\mu$m. The inset shows the amplification of optical/UV band for the sake of presentation.\label{component}}
	\vskip-10pt
\end{figure*}

\begin{figure*}
	\centering
\includegraphics[width=3.in]{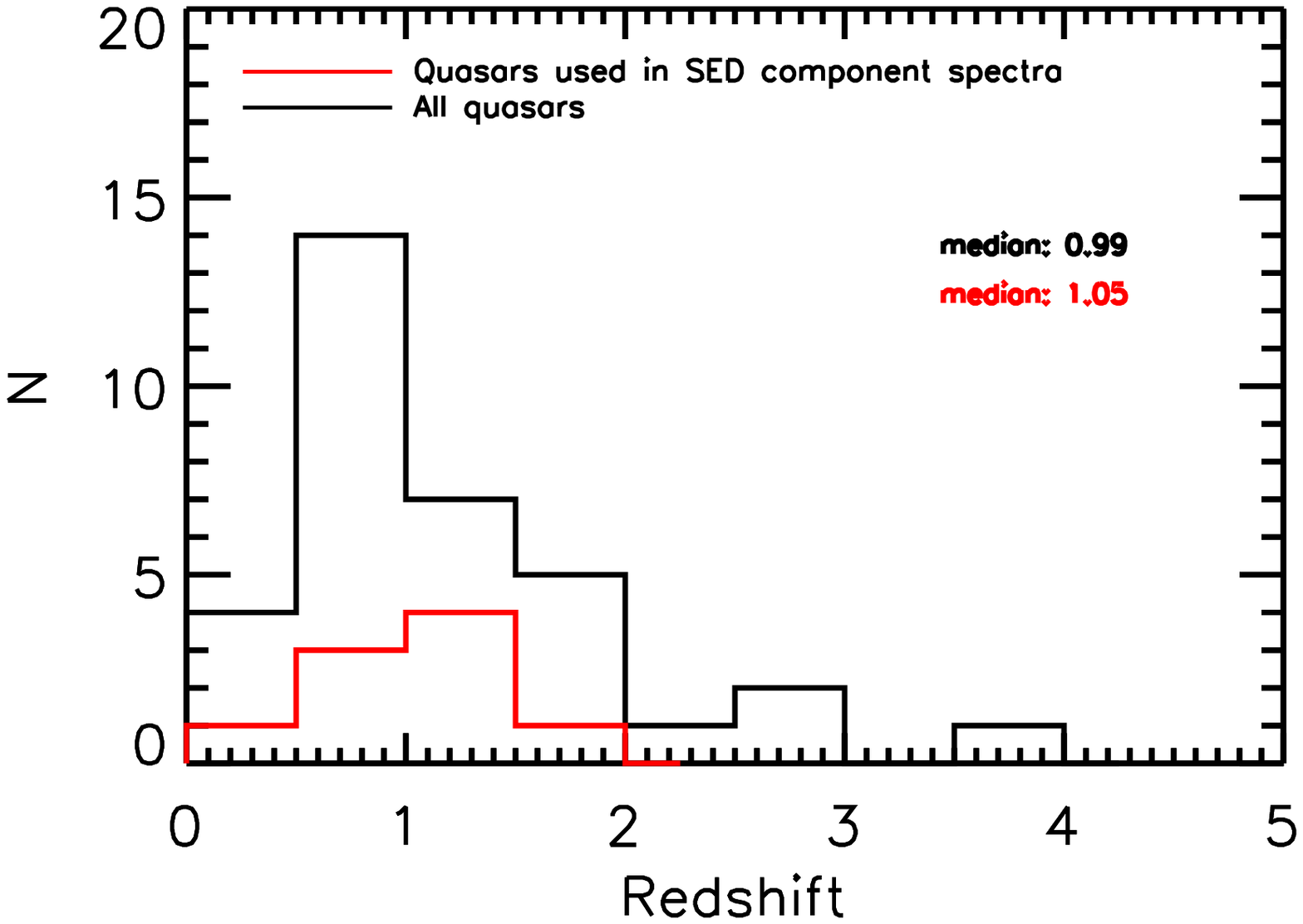}
\includegraphics[width=3.in]{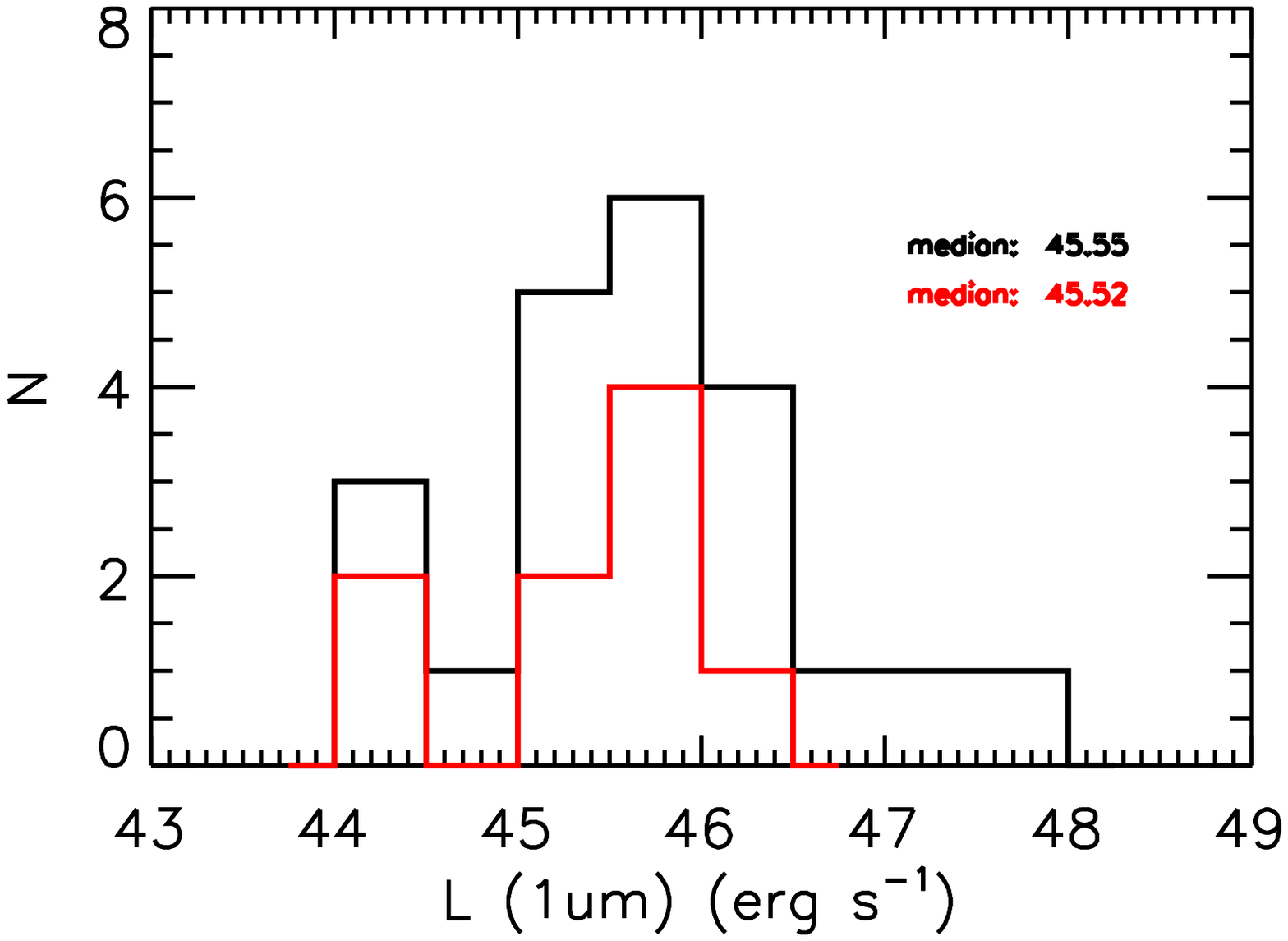}
\includegraphics[width=3.in]{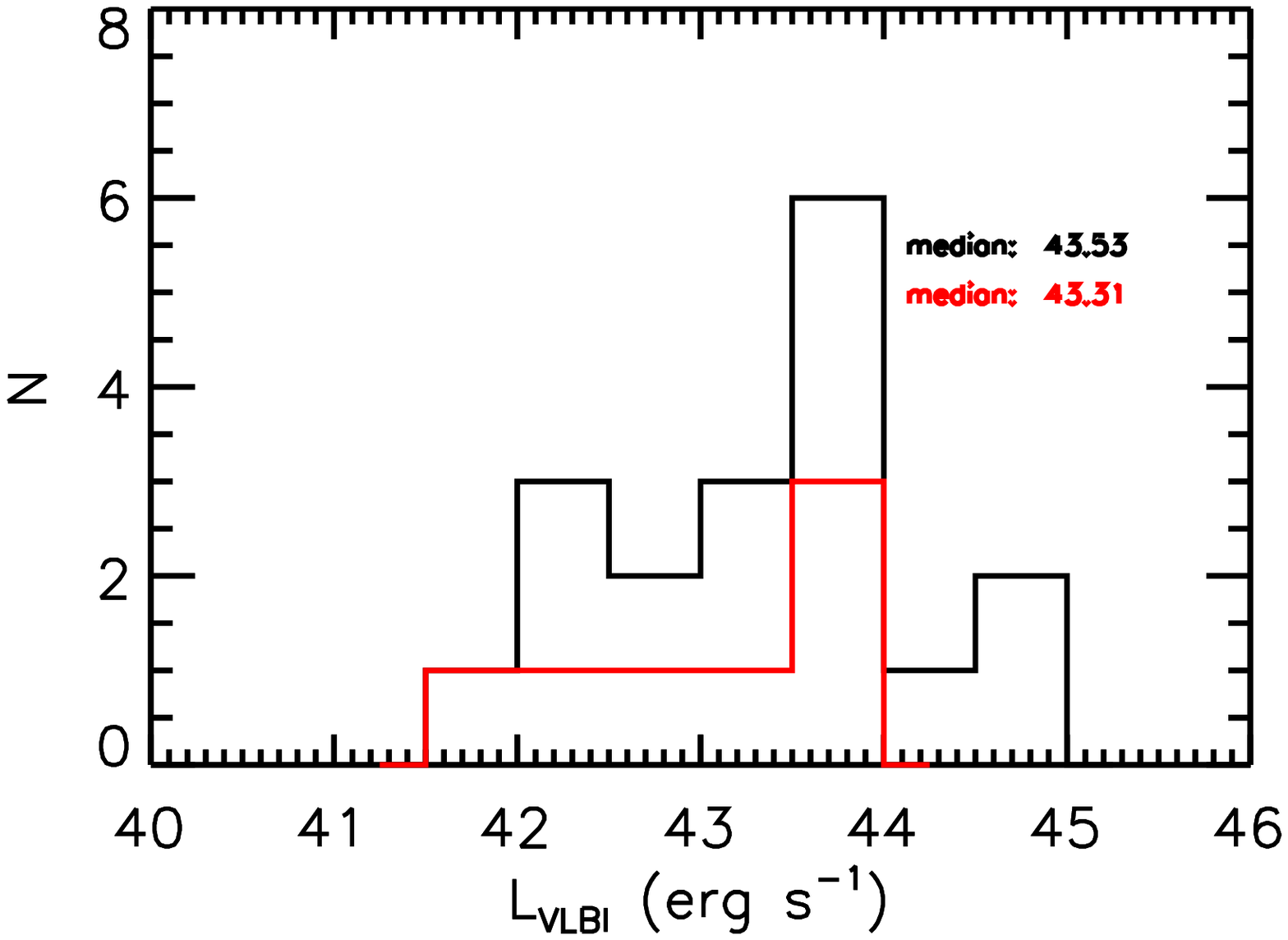}
\includegraphics[width=3.in]{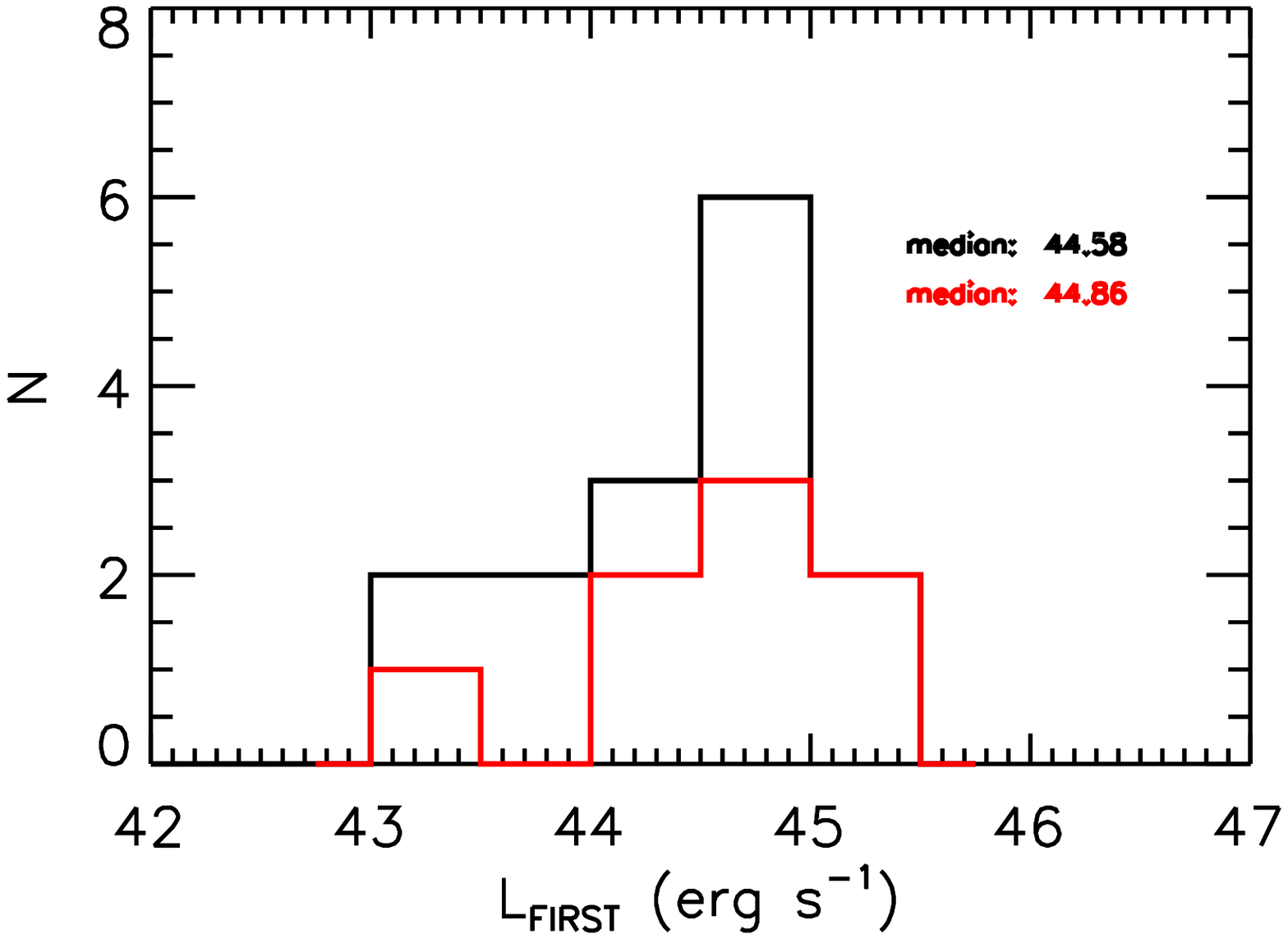}	
\includegraphics[width=3.in]{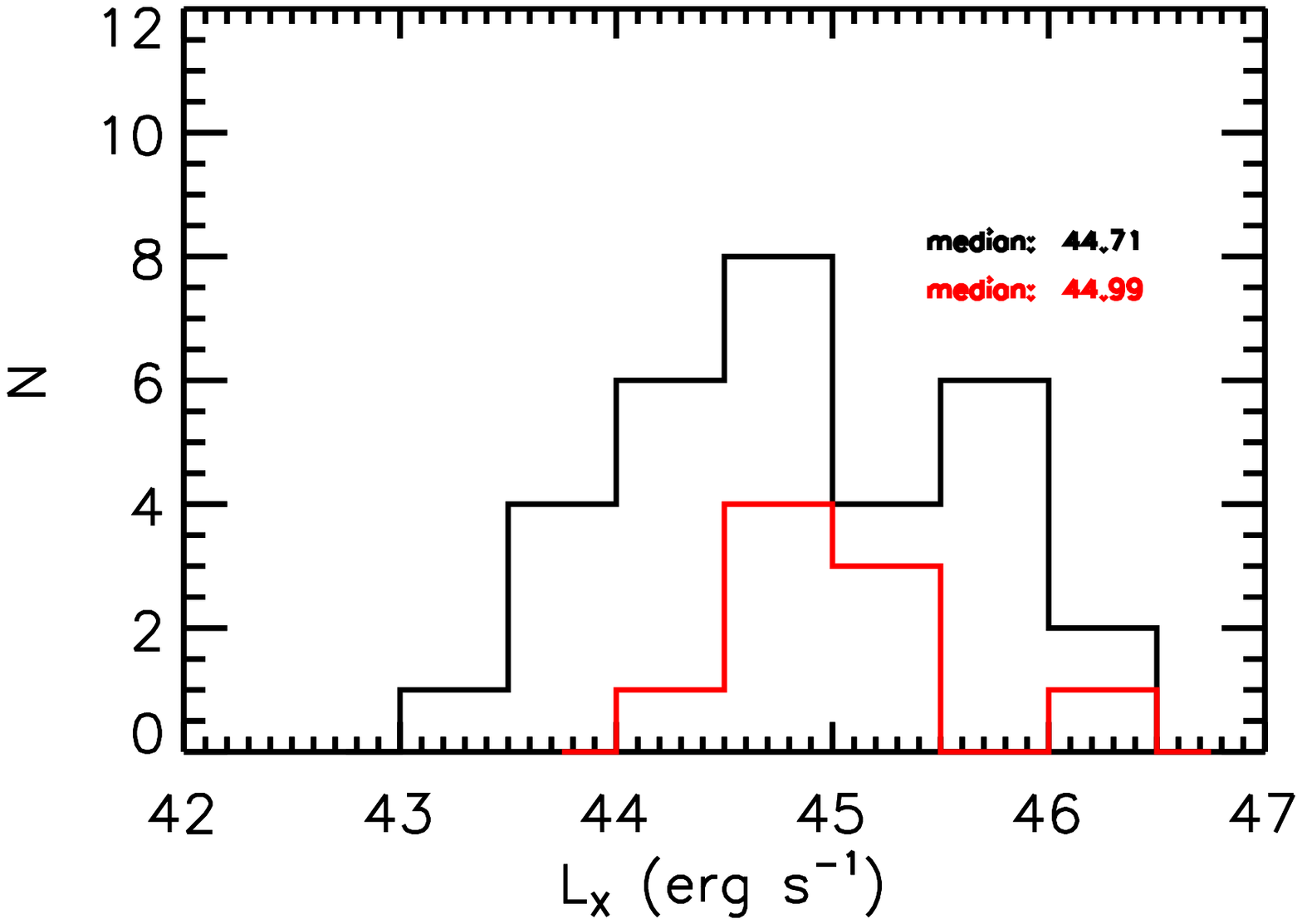}	
	\vspace*{-0.2 cm} \caption{The distributions of various quantities for all quasars in our sample (the black solid lines) and nine quasars used to compute the composite SED (the red solid lines): $top~ left:$ redshift; $top~ right:$ the luminosity of 1$\micron$ ; $middle~ left:$ the radio luminosity at 5 GHz with VLBI radio-core detection; $middle~ right:$ the radio luminosity at 5 GHz with FIRST detection; $bottom:$ the X-ray luminosity at 2$-$10 keV. \label{bias}}
	\vskip-10pt
\end{figure*}

We present the median composite SED in Figure \ref{component}. Since our composite SED was constructed based on only
nine quasars, it is necessary to study the properties of these nine quasars comparing to those of all quasars in our sample, and verify the representative of these nine quasars with respect to all quasars. The Kolmogorov$-$Smirnov (K$-$S) test was applied to check whether the differences of redshift, $ L_{1\micron}$, $L_{\rm R}$ and $L_{\rm X}$ distributions between nine quasars and all quasars are significant or not. The K-S test results show that the distributions are all similar for redshift, $L_{1\micron}$, $L_{\rm FIRST}$, $L_{\rm VLBI}$, and $L_{\rm X}$ (see Figure \ref{bias}), with P-value of 0.76, 0.29, and 0.99, 0.98 and 0.59, respectively. This implies that nine quasars can represent all quasars in our sample and there is likely no strong bias introduced to the composite SED, which is further supported by the consistency of the median values at 1.4 GHz and 6 keV between nine quasars and all quasars (indicated with blue crosses) as shown in Figure \ref{component}.

Although the composite SED is constructed from only nine quasars, it shows prominent features when compared with the composite SED of S11 quasars. While the overall SED at IR band is similar to S11 quasars \footnote{It should be noted that the two bins of 12.75-13.04 and 13.04-13.33 frequency range are only covered by three and two sources, respectively, implying large uncertainty in rebined value for these two bins to track the far-infrared emission.}, the radio spectrum of our composite SED is significantly larger than that of S11 RLQs, indicating powerful radio emission in our sources.
It can also be clearly seen that the hard X-ray emission in our comparison SED is higher than S11 RQQs though lower than S11 RLQs. With the normalization at $1 \micron$, the optical/UV spectrum of our composite SED is lower than that of S11 quasars, likely caused by dust extinction as argued in \cite{1995baker}.

\section{Discussion}
The theoretical models predict that the coefficients between the radio emission of optically thick core and X-ray emission are $\sim$ 0.6 or $\sim$ 1.4 when X-ray emission is from radiatively inefficient or efficient accretion flows, respectively \citep{2003MNRAS.345.1057M}, assuming a flat radio spectral index $\alpha_{\rm R}$ $\sim$ 0 and the canonical value of the power-law index of radio emitting electron $p$ = 2 \citep{2003MNRAS.343L..59H}. Therefore, if X-ray emission is from accretion flow in our high-accretion sources, $\rm b\sim 1.4$ would be expected. However, our results of b and $\xi_{\rm RX} \sim 1$ deviate from the predictions. Previous studies on the fundamental plane mainly focused on either in low-luminosity AGNs with radiatively inefficient accretion flows \citep[e.g.,][]{2003MNRAS.345.1057M}, or radiatively efficient accretion in radio-quiet sources \citep[e.g.,][]{2014ApJ...787L..20D}. Our young radio AGNs with both radiatively efficient accretion and powerful radio emission differ from these populations. The deviation of radio/X-ray correlation and fundamental plane from predictions indicate that the X-ray emission in these sources is not dominated by accretion process (i.e., disk-corona system), and the contribution from the jet might be important as they are strong radio emitters.

The jet-related X-ray emission could be associated with the synchrotron emission from the same population of electrons responsible for the radio emission, or inverse Compton (IC) scattering of these electrons on low-energy photons. The IC emission could be either synchrotron self-Compton (SSC) where the seed photons are produced in the jet, or the inverse Compton on an external seed photon field (EC) \citep{1992A&A...256L..27D}. As suggested in \cite{2006A&A...456..439K}, the cut-off photon energy of synchrotron emission is usually below X-ray band in strongly accreting system.
The average value of photon index at 2$-$10 keV is 1.64 for the sources involved in the fundamental plane.
The value is similar to that of flat spectrum radio quasars (FSRQs, 1.65$\pm 0.04$) \citep{2001A&A...375..739D}, where the hard X-ray emission is general due to SSC \citep{2018ApJS..235...39C}. On the other hand, the photon index of our sources is flatter than normal RQQs \cite[$\Gamma\sim1.89$,][]{2006A&A...456..439K}. Moreover, no significant correlation was found between hard X-ray photon index and Eddington ratio (see Section 3.2.1). In contrast, the correlation was found to be very strong in radio-quiet quasars of which the X-ray is produced by disk-corona system \citep{2004ApJ...607L.107W,2019MNRAS.490.3793L}. All these results imply that the X-ray emission in our sources could be different from RQQs, and may be dominated by the jet emission, presumably through SSC. 
The notion that the X-ray emission is likely from SSC is supported by the comparison of VLBI radio-core flux and X-ray flux for the sample of 56 extragalactic sources in \cite{1991ApJ...366...16B}, which suggested their data are consistent with the expections of SSC model. The analysis on their data shows $\rm b=1.06 \pm 0.08$ ($L_{\rm R}\propto L^{\rm b}_{\rm X}$) by using the same OLS method as in our work, in good agreement with our results.

Although the blazar-type objects have been excluded in our original sample \citep{paper1}, our median composite SED of quasars have clear excess at hard X-ray band with respect to the composite SED of S11 RQQs. This may also be the clue of jet-related emission at X-ray band. 
The broadband SED modeling of young radio AGNs have been made by several studies, which can be effective in discerning the X-ray emission components and possible jet contribution. In \cite{2012ApJ...749..107M}, they showed that the jet of CSS quasar 3C 186 could contribute to the total X-ray emission when it develops a complex velocity structure. 
In another CSS quasar, 3C 48, the excess in the hard X-ray band reported in \cite{2004MNRAS.347..632W} is ascribed to the inner part of the jet. \cite{2014ApJ...780..165M} used synchrotron+IC model to fit the SEDs of a sample of young quasars from \cite{2008ApJ...684..811S}, most of which is also included in our sample. The authors claimed that the bulk of X-ray emission should be related to the inner jet ($<$ 1 kpc) if it is from jet. This is in line with our finding that the X-ray emission is tightly related with the VLBI core emission commonly within 1 kpc. The SED modeling of individual sources in a large sample is necessary to further study the mechanism of X-ray emission, which however is beyond the scope of our present work.

\begin{figure*}
	\centering
	\includegraphics[width=3.0in]{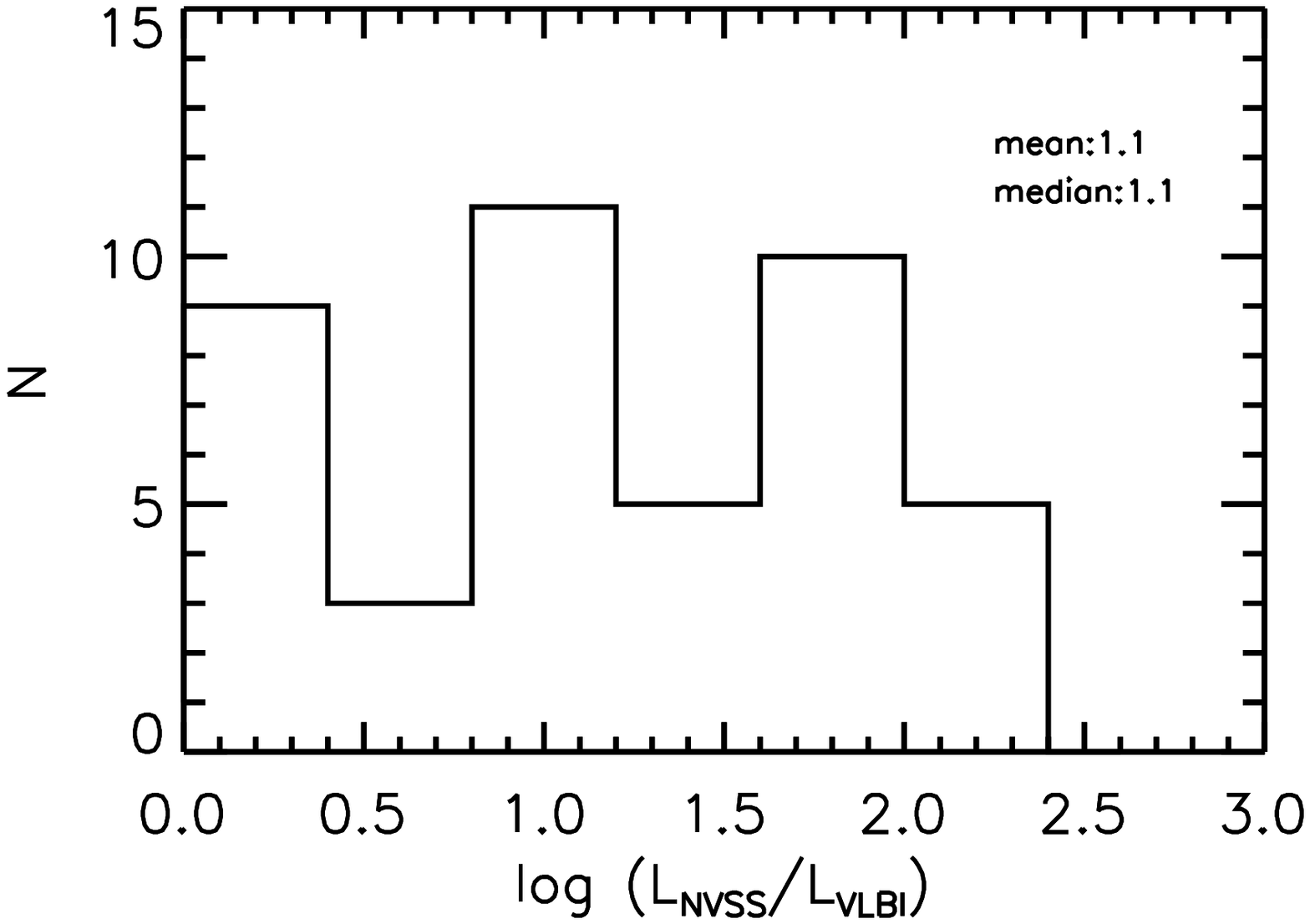}	
    \includegraphics[width=3.0in]{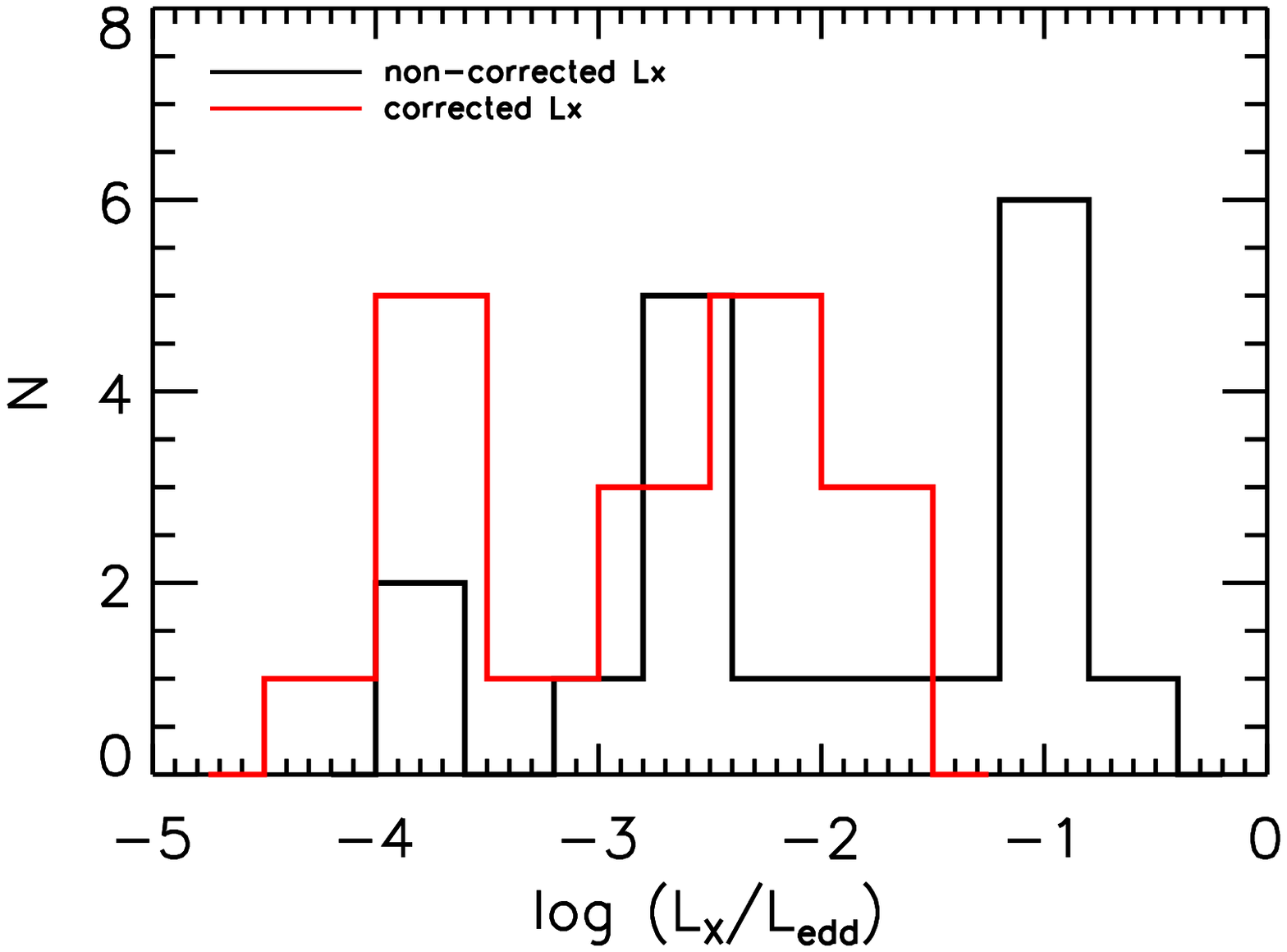}
	\includegraphics[width=3.0in]{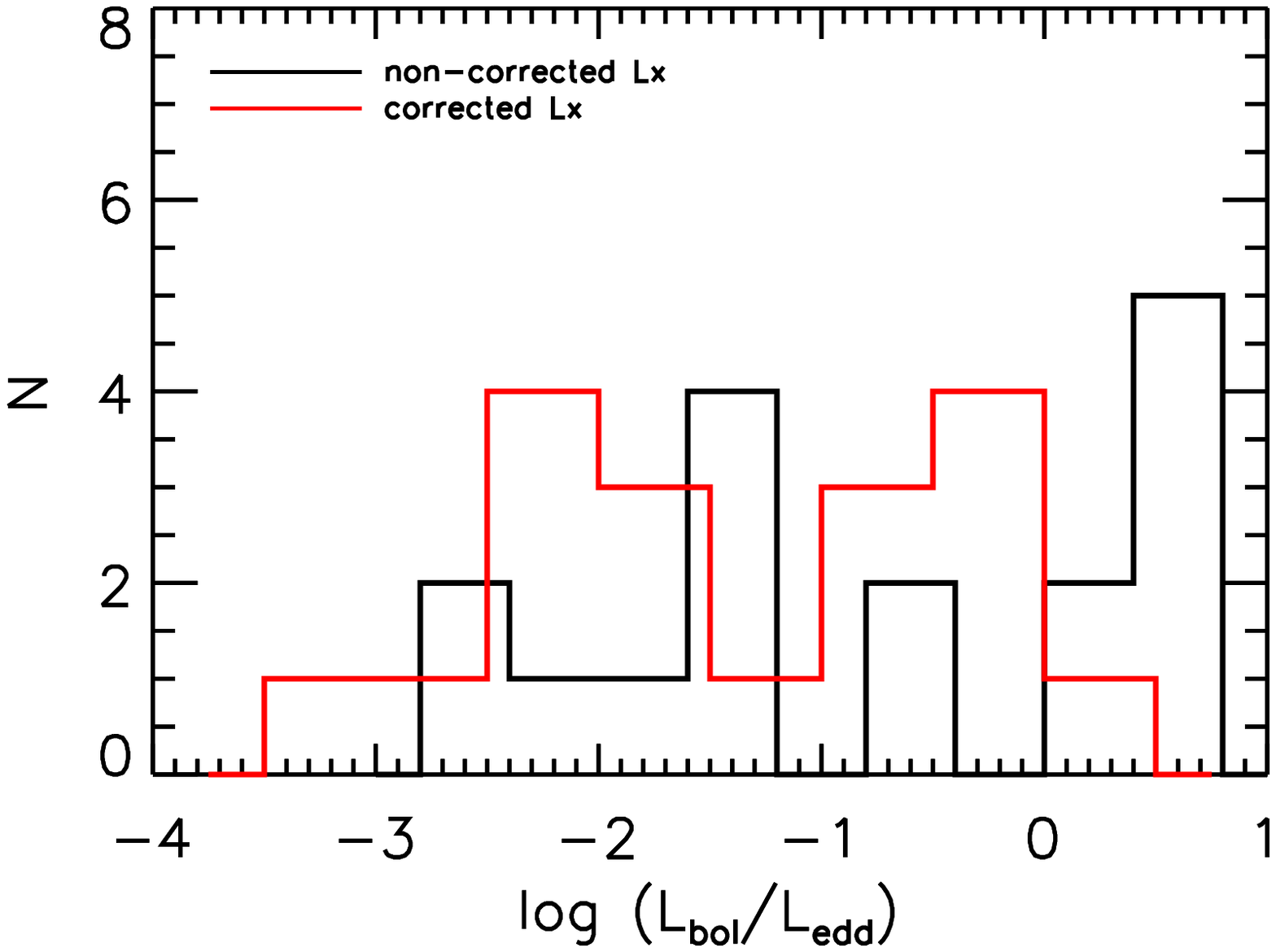}
	\includegraphics[width=3.0in]{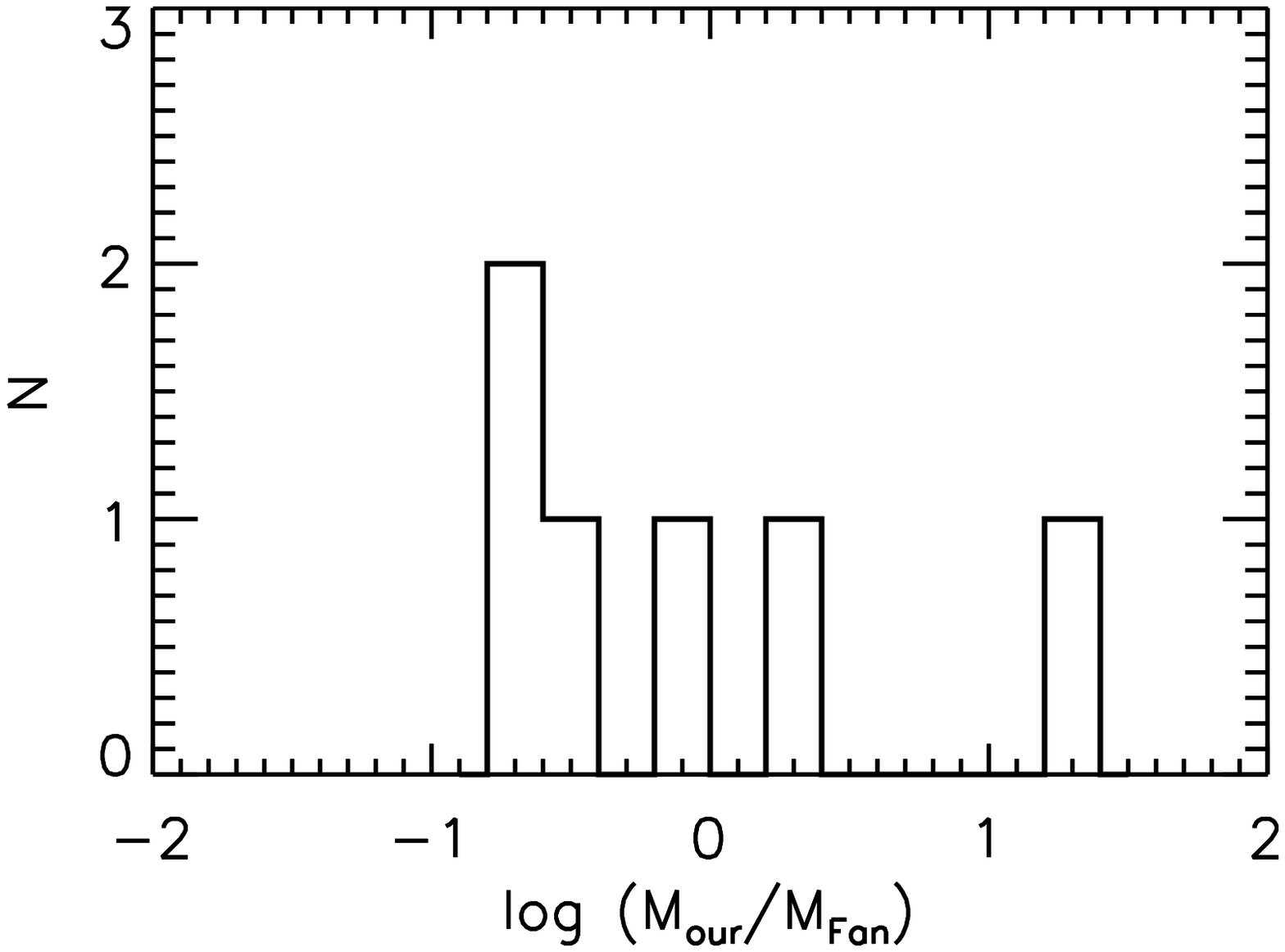}
	\vspace*{-0.2 cm}\caption{$upper~ left:$ the radio luminosity ratio for our sources with both VLBI and NVSS detections; $upper~ right:$ the distributions of $L_{\rm X}/L_{\rm Edd}$ for 18 sources in the FP study in Fan $\&$ Bai (2016) before and after correcting $L_{\rm X}$; $lower~ left:$ the distributions of $L_{\rm bol}/L_{\rm Edd}$ for 18 sources in the FP study in Fan $\&$ Bai (2016) before and after correcting $L_{\rm X}$; $lower~ right:$ the difference on BH mass for six common sources between our work and Fan $\&$ Bai (2016). \label{nvss_first}}
	\vskip-10pt
\end{figure*}

The first systematic study of radio/X-ray relation and fundamental plane for young radio AGNs is shown in \cite{2016ApJ...818..185F}. They found that radio/X-ray connection b ($L_{\rm R}$ $\propto$ $L_{\rm X} ^{b}$) and $\xi_{\rm RX}$ $\sim$ 0.6, consistent with the result of \cite{2003MNRAS.345.1057M}, which support the accretion-related X-ray emission. Our finding of b and $\xi_{\rm RX}$ $\sim$ 1 in our high-accretion sources ($R_{\rm edd} \ga 10^{-3}$) is apparently different from those of \cite{2016ApJ...818..185F}.
Although both works are based on the comparable sample size in studying the fundamental plane (see Table 2), our sample requires the sources with the higher-resolution VLBI detections which can effectively exclude the dominant large-scale radio emission (e.g., in NVSS images), and with uniformly estimated black hole masses. In order to investigate the reasons for the steepening $\xi_{\rm RX}$, we started from checking the data of 31 commom sources between our work and \cite{2016ApJ...818..185F} (see Table 2). All based on the original data from \cite{2008ApJ...684..811S} and \cite{teng09}, we found that the $L_{\rm X}$ in their work were overestimated by about one order of magnitude compared to ours (confirmed by private communication with the authors). With the corrected $L_{\rm X}$, their fundamental plane (FP) relation was then re-analyzed. We found that the new index is $\xi_{\rm RX} = 0.66 \pm$0.08, still consistent within the errors with their original result, 0.58$\pm$0.03. Since the corrected $L_{\rm X}$ is lower than the original values, we then studied the FP only for $L_{\rm bol}/ L_{\rm edd}$ $>$ $10^{-3}$ (as in our work) by converting the corrected $L_{\rm X}$ to $L_{\rm bol}$ by applying the bolometric correction at X-ray band from \cite{2019MNRAS.488.5185N}. In this case, only one source in their sample, PKS 0941-08 ($L_{\rm bol}/L_{\rm edd}$ $< 10^{-3}$) was excluded (see Figure \ref{nvss_first}), which has the least $L_{\rm X}$ and locate in the low luminosity end of $L_{\rm R}$. Interestingly, the derived $\xi_{\rm RX}$ value (0.92$\pm$0.05) is consistent with our result (0.99$\pm$0.05) within errors (see Table 3). This could give us the hint of importance of FP study at different range of accrete rate.
The consistent result while using different radio data can be understood with the consistent correlations from VLA, FIRST and VLBI, and especially $L_{\rm VLA/FIRST} ~\sim$ 10 $L_{\rm VLBI}$. We analyzed those 44 sources in our whole sample with both VLBI and NVSS detections and found that averagely $L_{\rm NVSS} ~\sim$ 10 $L_{\rm VLBI}$ (albeit large scatter, see Figure \ref{nvss_first}), similar to $L_{\rm FIRST} ~\sim$ 10 $L_{\rm VLBI}$. 
This supports that the large scale radio emission probed by NVSS will not significantly alter $\xi_{\rm RX}$. In constructing our sample, we required the uniformly estimated black hole masses based on SDSS spectra, in contrast to the collection from the literature in \cite{2016ApJ...818..185F}. The discrepancy of BH masses in six overlapped sources in the FP study between our work and \cite{2016ApJ...818..185F} is shown in Figure \ref{nvss_first} (see also Table 2). We found that the discrepancies in all sources are below or close to the typical uncertainty 0.5 dex in BH mass estimation, except for one object with 1.33 dex, in which the BH mass calculated from stellar velocity dispersion is adopted by us, while the value obtained based on broad emission lines was used in \cite{2016ApJ...818..185F}. We found that the replacement of BH masses with our measurements in these six objects still gives consistent $\xi_{\rm RX}$ with ours (see Table 3). Our re-analysis on Fan \& Bai's sample indicates that  the results in our work is generally consistent with Fan \& Bai at $L_{\rm bol}/ L_{\rm edd}$ $>$ $10^{-3}$, although different radio emission are used (VLBI versus NVSS) and only six sources are overlapped in two samples.

It should be noted that the number of sources included in the studies on the fundamental plane and spectral energy distribution is rather small. Moreover, the similar analysis can't be done for sources with low Eddington ratio (i,e., radiatively inefficient accretion) due to the limited data and source number. 
A larger sample with better observational data will be needed to further study the mechanism of X-ray emission for young radio AGNs, at both high and low accretion rates. 

\section{Conclusion}

In this work, we study the X-ray emission in young radio AGNs by using multi-band data. We found that the radio and X-ray luminosities are tightly correlated with linear relations. The fundamental plane of high-accreted sources are investigated, and the approximate linear dependence of radio on X-ray emission is found. The photon index of hard X-ray is similar to that of FSRQs, and there is no significant correlation between the photon index and Eddington ratio. The X-ray spectra in most individual objects and composite SED for young radio quasars are higher than that of RQQs. All our results imply that the X-ray emission in high-accreted young radio AGNs could be related with jet, likely from SSC.

\section*{Acknowledgements}
We thank the anonymous referee for valuable comments and useful suggestions to improve the manuscript.
We thank Xuliang Fan and Shuangliang Li for useful discussions.
This work is supported by the National Science Foundation of China (grants 11873073, U1531245, 11773056, 11473054 and U1831138).
Funding for SDSS-III has been provided by the Alfred P. Sloan Foundation, the Participating Institutions, the National Science Foundation, and the U.S. Department of Energy Office of Science. The SDSS-III web site is http://www.sdss3.org/. 
This research has made use of the NASA/IPAC Extragalactic Database (NED) which is operated by the Jet Propulsion Laboratory, California Institute of Technology, under contract with the National Aeronautics and Space Administration. This research has made use of data obtained from the Chandra Data Archive and the Chandra Source Catalog, and software provided by the
Chandra X-ray Center (CXC) in the application packages CIAO, ChIPS, and Sherpa. This work is based on observations obtained with XMM-Newton, an ESA science mission with instruments and contributions directly funded by ESA Member States and NASA.

\begin{table*}
	\caption{Sample}
	\label{tab:landscape}
	\setlength{\tabcolsep}{0.04in}
	\begin{center}
		\begin{tabular}{lcccrccrccrcccccc}
			\hline \hline
			Name & $z$ & Type & ID & log $F_{\rm R}$ & log $L_{\rm R}$ & refs. & log $F_{\rm R}$ & log $L_{\rm R}$ & refs. &  log $F_{\rm R}$ & log $L_{\rm R}$ & log $L_{\rm X}$ & $\Gamma$ & refs. &log $M_{\rm BH}$ &  log $R_{\rm edd}$ \\
			&  &   &   & VLA  &   &  & VLBI &   &  & FIRST &  & 2-10keV & &  &  & \\
			(1) & (2) & (3)  & (4)  & (5)  &  (6) & (7) & (8) & (9)  & (10) &(11) &(12) &(13) &(14) & (15) & (16) & (17) \\
			\hline
			3C 43 & 1.459 & CSS   & Q     & 950   & 44.41  & 1     & 175   & 43.67 & 17    & . . . & . . . & 45.57 & $1.53^{+0.25}_{-0.15}$ & 42    & . . . & . . .\\
			3C 48 & 0.367 & CSS   & Q     & 5700  & 44.00 & 1     & 61 & 42.01   & 18    & . . . & . . . & 44.70  & $2.52^{+0.07}_{-0.09}$  & 43    & . . . & . . . \\
			3C 49$^{\spadesuit}$ & 0.621 & CSS   & Q     & 660   & 42.75 & 2     & . . . & . . . & . . . & . . . & . . . & 44.71 & $1.66^{+0.33}_{-0.32}$  & 42    & 8.59 & -2.05 \\
			3C 67 & 0.310  & CSS   & G     & . . . & . . . & . . . & 20 & 41.43 &  19 & . . . & . . . & 43.99 & $1.32^{+0.13}_{-0.11}$  & 44    & . . . & . . . \\
			3C 93.1 & 0.243 & CSS   & G     & 790   & 42.75  & 1     & . . . & . . . &  . . . & . . . & . . . & 43.24 & $1.87^{+0.31}_{-0.29}$  & 42    & . . . & . . . \\
			3C 119 & 1.023 & CSS   & G    & 3750  & 44.71 & 1     & 95    & 43.11 &  20    & . . . & . . . & 45.08 & $1.50^{+0.10}_{-0.10}$   & 45    & . . . & . . . \\
			3C 138 & 0.759 & CSS   & Q     & 1089  & 43.91 & 3     & 38    & 42.46 &  21    & . . . & . . . & 45.66 & $1.46^{+0.24}_{-0.11}$  & 42    & . . . & . . . \\
			3C 147 & 0.545 & CSS   & Q     & 8000  & 44.49 & 1     & . . . & . . . &  . . . & . . . & . . . & 44.86 & $1.85^{+0.24}_{-0.22}$  & 42    & . . . & . . . \\
			3C 173 & 1.035 & CSS   & G     & 373   & 43.72 & 4     & . . . & . . . & . . . & . . . & . . . & 44.43 & 1.70$^a$   & 45    & . . . & . . . \\
			3C 186$^{\clubsuit}$ $^{\spadesuit}$ & 1.066 & CSS   & Q     & . . . & . . . & . . . & 18    & 42.43 &  19    & 1245 & 44.15 & 44.64 & $2.09^{+0.08}_{-0.08}$  & 43    & 9.11 & -1.26 \\
			3C 190$^{\clubsuit}$ $^{\spadesuit}$ & 1.196 & CSS   & Q     & 100   & 43.27 & 1     & 21$^{\ast}$    & 42.59 &  22    & 2598 & 44.58 & 45.26 & $1.61^{+0.06}_{-0.05}$  & 42    & 8.82 & -1.33 \\
			3C 191$^{\clubsuit}$ & 1.968 & CSS   & Q     & 35    & 43.21 & 1     & 22$^{\ast}$  & 43.00 &  22    & 1870 & 44.90 & 45.65 & $1.68^{+0.11}_{-0.10}$  & 42    & 8.06 & 0.25 \\
			4C 40.20 & 0.551 & CSS   & Q     & . . . & . . . & . . . & . . . & . . . &  . . . & 1068 & 43.44 & 44.23 & $1.91^{+0.37}_{-0.36}$  & 46    & 8.05 & -1.25 \\
			4C 07.22 & 0.112 & CSS   & G     & . . . & . . . & . . . & . . . & . . . &  . . . & 463 & 41.57 & 42.80 & $0.64^{+0.15}_{-0.15}$  & 47    & 8.37 & -2.93 \\
			3C 213.1$^{\clubsuit}$ & 0.194 & CSS   & G     & 460   & 42.32 & 4     & 16$^{\ast}$    & 40.86 &  19    & 1493 & 42.59 & 42.68 & $1.90^{+0.44}_{-0.36}$   & 42    & 7.83 & -2.20 \\
			3C 216$^{\spadesuit}$ & 0.670  & CSS   & Q     & 1050  & 43.79  & 1     & 350   & 43.31 &  23    & 4010 & 44.20 & 45.15 & $1.59^{+0.26}_{-0.13}$  & 42    & . . . & . . . \\
			SDSS J090951.11+044422.7 & 0.640  & CSS   & G     & . . . & . . . & . . . & . . . &  . . . & . . . & 185 & 42.82 & 42.90 & 1.70  & 47    & . . . & . . . \\
			SDSS J092405.30+141021.4 & 0.136 & CSS   & G     & . . . & . . . & . . . & . . . & . . . & . . . & 108 & 41.12 & 42.18 & $1.80^{+0.30}_{-0.30}$   & 48    & 9.17 & -3.83 \\
			SDSS J094525.90+352103.6 & 0.208 & CSS   & G     & . . . & . . . & . . . & . . . &  . . . & . . . & 148 & 41.65 & 42.91 & $1.60^{+0.24}_{-0.16}$  & 47    & 7.70 & -1.37 \\
			3C 237 & 0.877 & CSS   & Q     & . . . & . . . & . . . & . . . & . . . &  . . . & 6401 & 44.67 & 43.91 & $1.88^{+0.44}_{-0.40}$  & 42    & 9.84 & -2.66 \\
			3C 241 & 1.617 & CSS   & G     & . . . & . . . & . . . & . . . & . . . &  . . . & 1734 & 44.67 & 45.00    & $1.47^{+0.25}_{-0.24}$  & 42    & . . .  & . . . \\
			4C 35.23     & 1.604 & CSS   & Q     & . . . & . . . & . . . & 104 & 43.53 &  24    & 1051 & 44.46 & 43.84 & 1.70   & 49    & . . . & . . . \\
			3C 258 & 0.165 & CSS   & G     & . . . & . . . & . . . & . . . & . . . &  . . . & 897 & 42.22 & 41.72 & 1.70$^a$    & 50    & . . . & . . . \\
			4C 46.23 & 0.116 & CSS   & G     & 149 & 41.37 & 5     & . . . & . . . & . . . & 423 & 41.57 & 41.40  & 1.90   & 51    & 8.89 & . . . \\
			3C 268.3 & 0.372 & CSS   & Q     & . . . & . . . & . . . & . . . & . . . &  . . . & . . . & . . . & 44.34 & $1.85^{+0.60}_{-0.50}$  & 6     & 8.24 & -1.97 \\
			3C 275 & 0.480  & CSS   & G     & . . . & . . . & . . . & . . . & . . . &  . . . & 3600 & 43.83 & 43.50  & 1.70$^a$  & 44    & . . . & . . . \\
			3C 277.1$^{\clubsuit}$ $^{\spadesuit}$ & 0.320  & CSS   & Q     & 500   & 42.80 & 1     & 26     & 41.58 &  19    & 2442 & 43.27 & 44.18 & $1.85^{+0.07}_{-0.07}$  & 43    & 8.37 & -1.44 \\
			SDSS J132419.67+041907.0 & 0.263 & CSS   & G     & 14    & 41.07 & . . . &  . . . & . . . & . . . & 155 & 41.89 & 42.31 & $2.35^{+0.39}_{-0.36}$  & 47    & . . . & . . . \\
			3C 287$^{\clubsuit}$ $^{\spadesuit}$ & 1.055 & CSS   & Q     & 3280  & 44.68 & 1     & 560   & 43.91 &  25    & 6999 & 44.89 & 44.99 & $1.86^{+0.07}_{-0.05}$  & 43    & 8.38 & -0.73 \\
			3C 286$^{\clubsuit}$ $^{\spadesuit}$ & 0.850  & CSS   & Q     & 6100  & 44.76 & 1     & 380   & 43.56 &  26    & 15026 & 45.01 & 44.95 & $2.12^{+0.51}_{-0.26}$  & 42    & 8.14 & -0.70 \\
			3C 298$^{\clubsuit}$ $^{\spadesuit}$ & 1.436 & CSS   & Q     & 1700  & 44.65 & 3     & 330   & 43.94 & 17    & 6156 & 45.13 & 46.08 & $1.85^{+0.05}_{-0.05}$  & 43    & 8.95 & -0.69 \\
			PKS B1421-490 & 0.662 & CSS   & G     & . . . & . . . & . . . & . . . & . . . & . . . & . . . & . . . & 44.21 & $1.26^{+0.14}_{-0.13}$  & 42    & . . . & . . . \\
			3C 303.1 & 0.270  & CSS   & G     & . . . & . . . & . . . & . . . & . . .  & . . . & . . . & . . . & 42.46 & 1.70$^a$  & 6     & . . .& . . . \\
			3C 305.1 & 1.132 & CSS   & G     & . . . & . . . & . . . & . . . & . . .  & . . . & . . . & . . . & 44.19 & 1.70$^a$ & 45    & . . . & . . . \\
			B3 1445+410 & 0.195 & CSS   & G     & 26    & 41.07 & 6     & . . . & . . . &  . . . & 397 & 42.02 & 42.82 & $1.67^{+0.11}_{-0.11}$  & 6     & 8.81 & -3.59 \\
			3C 305 & 0.042 & CSS   & G     & . . . & . . . & . . . & . . . & . . . &  . . . & 2947 & 41.50 & 40.70  & $1.70^{+0.05}_{-0.05}$   & 53    & 8.02 & -2.47 \\
			3C 309.1 & 0.905 & CSS   & Q     & 804   & 43.94 & 3     & 287 & 43.49 &  27    & . . . & . . . & 45.55 & $1.57^{+0.04}_{-0.04}$  & 43    & . . . & . . . \\
			3C 318$^{\spadesuit}$ & 1.572 & CSS   & Q     & 500   & 44.19  & 1     & . . . & . . . &  . . . & 2779 & 44.86 & 45.28 & $1.88^{+0.20}_{-0.19}$  & 42    & 8.75 & -1.40 \\
			B2 1542+39   & 0.553 & CSS   & G     & . . . & . . . & . . . & . . . &  . . . & . . . & 195 & 42.98 & 42.70  & 1.70    & 47    & 8.76 & -2.48 \\
			SDSS J155927.67+533054.4 & 0.179 & CSS   & G     & . . . & . . . & . . . &  . . . & . . . & . . . & 182 & 41.60 & 41.71 & 1.70   & 47    & 8.34 & -2.74 \\
			4C 52.37 & 0.106 & CSS   & G     & . . . & . . . & . . . & 38  & 40.69 &  28    & 576 & 41.62 & 41.70  & $1.34^{+1.10}_{-0.47}$  & 54    & 8.69 & -3.61 \\
			PMN J1626+0448 & 0.040  & CSS   & G     & . . . & . . . & . . . & . . . &  . . . & . . . & 186 & 40.25 & 40.01 & 1.70   & 47    & . . . & . . . \\
			3C 343 & 0.988 & CSS   & Q     & 1460  & 44.27 & 1     & . . . & . . . & . . . & 4988 & 44.68 & 43.57 & 1.7$^a$  & 55    & . . .  & . . . \\
			3C 343.1 & 0.750  & CSS   & G     & 1250  & 43.96 & 1     & . . . & . . . &  . . . & 4741 & 44.39 & 44.04 & $1.51^{+1.09}_{-0.90}$  & 42    & . . . & . . . \\
			3C 346 & 0.163 & CSS   & G     & . . . & . . . & . . . & . . . & . . . &  . . . & 3675 & 42.82 & 43.45 & $1.74^{+0.05}_{-0.05}$  & 42  & 8.67 & -3.36 \\
			B3 1702+457 & 0.060  & CSS   & G     & . . . & . . . & . . . & 56  & 40.37 &  29    & 119 & 40.43 & 43.76 & 2.27  & 56    & . . . & . . . \\
			4C 13.66 & 1.450  & CSS   & G     & . . . & . . . & . . . & . . . & . . . & . . . & . . . & . . . & 43.60  & 1.90 & 57   & . . . & . . . \\
			3C 380 & 0.691 & CSS   & Q     & 2800  & 44.24 & 1     & 1039 & 43.81 &  27    & . . . & . . . & 45.81 & $1.54^{+0.09}_{-0.09}$  & 58    & . . .  & . . . \\
			4C 29.56 & 0.842 & CSS   & Q     & . . . & . . . & . . . & . . . &  . . . & . . . & . . . & . . . & 43.49 & $1.82^{+0.72}_{-0.66}$  & 43    & . . . & . . . \\
			3C 454 & 1.757 & CSS   & Q     & . . . & . . . & . . . & . . . &  . . . & . . . & . . . & . . . & 45.52 & $1.68^{+0.09}_{-0.07}$  &  59   & . . .  & . . . \\
			3C 455 & 0.543 & CSS   & Q     & . . . & . . . & . . . & . . . & . . . & . . . & . . . & . . . & 44.08 & $1.65^{+0.12}_{-0.21}$  & 42    & . . .  & . . .\\
			4C 31.04 & 0.059 & CSS   & G     & . . . & . . . & . . . & 22    & 39.95   & 30    & . . . & . . . & . . . & . . . & . . .    & . . . & . . . \\
			4C 00.02$^{\clubsuit}$ & 0.306 & GPS   & G     & 1100  & 43.10 & 7     & 47$\ast$ & 41.79 &  19 & 2919 & 43.31 & 42.73 & . . .$^b$ & 60    & 8.55 & -2.96 \\
			PKS 0237-23 & 2.225 & GPS   & Q     & 3340  & 45.29 & 7     & 290 & 44.48 &  19 & . . . & . . . & 44.66 & $1.72^{+0.03}_{-0.03}$  & 42    & . . .  & . . . \\
			4C 05.19   & 2.639 & GPS   & Q     & . . . & . . . & . . . & . . . &  . . . & . . . & . . . & . . . & . . . & . . .  & . . .    & . . . & . . . \\
			PKS 0428+20 & 0.219 & GPS   & G     & 2300  & 43.12 & 7     & 29    & 41.22 &  32    & . . . & . . . & 43.14 & . . .$^b$ & 60     & . . . & . . . \\
			4C 14.41 & 0.363 & GPS   & G     & 1000  & 43.22 & 7     & . . . & . . . &  . . . & 2439 & 43.40 & 43.14 & . . .$^b$ & 60    & 9.10 & -3.13 \\
			PKS 1127-14 & 1.184 & GPS   & Q     & 3820  & 44.84 & 7     & 1890  & 44.54 &  23    & . . . & . . . & 46.24 & $1.20^{+0.03}_{-0.03}$   & 43    & . . . & . . . \\
			PKS 1143-245 & 1.950  & GPS   & Q     & 1400  & 44.81 & 7     & 1100  & 44.70  & 33    & . . . & . . . & 45.29 & $1.62^{+0.26}_{-0.22}$  & 43    & . . . & . . . \\
			PKS 1245-19 & 1.280  & GPS   & Q     & 2300  & 44.69 & 7     & . . . & . . . &  . . . & . . . & . . . & 44.09 &$1.96^{+0.45}_{-0.43}$  & 43    & . . . & . . . \\
			\hline
		\end{tabular}
	\end{center}
\end{table*}
\addtocounter{table}{-1}

\begin{table*}
	\setlength{\tabcolsep}{0.04in} \caption{Continued\dots
		\label{Sample}}
	\begin{center}
		\begin{tabular}{lcccrccrccrcccccc}
			\hline \hline
			Name & $z$ & Type & ID & log $F_{\rm R}$ & log $L_{\rm R}$ & refs. & log $F_{\rm R}$ & log $L_{\rm R}$ & refs. &  log $F_{\rm R}$ & log $L_{\rm R}$ & log $L_{\rm X}$ & $\Gamma$ & refs. &log $M_{\rm BH}$ &  log $R_{\rm edd}$ \\
			&  &   &   & VLA  &   &  & VLBI &   &  & FIRST &  & 2-10keV & &  &  & \\
			(1) & (2) & (3)  & (4)  & (5)  &  (6) & (7) & (8) & (9)  & (10) &(11) &(12) &(13) &(14) & (15) & (16) & (17) \\
			\hline
			B3 1315+415 & 0.066 & GPS   & G     & . . . & . . . & . . . & 127 & 40.80 &  28    & 249 & 40.83 & 42.06 & 1.70   & 61    & 8.84 & -4.30 \\
			4C 32.44 & 0.370  & GPS   & G     & 2350  & 43.61 & 7     & . . . & . . . & . . . & 4747 & 43.70 & 43.56 & $1.74^{+0.20}_{-0.20}$  & 60    &  &  \\
			4C 12.50$^{\clubsuit}$ & 0.122 & GPS   & G     & 3030  & 42.72 & 7     & 440   & 41.88   & 33    & 4860 & 42.67 & 43.60  & $1.10^{+0.29}_{-0.28}$   & 43    & 8.63 & -2.93 \\
			PKS 1607+26 & 0.474 & GPS   & G     & 1710  & 43.69 & 7     & . . . & . . . & . . . & 4845 & 43.95 & 43.50  & $1.40^{+0.10}_{-0.10}$   & 62    & 9.03 & -2.57 \\
			2MASX J19455354+7055488 & 0.101 & GPS   & G     & . . . & . . . &  . . . & . . . & . . . & . . . & . . . & . . . & 43.08 & $1.10^{+0.30}_{-0.10}$   & 63     & . . .  & . . . \\
			PKS 2008-068 & 0.547 & GPS   & G     & 1340  & 43.71 & 7     & 50    & 42.28 & 34    & . . . & . . . & 43.77 & . . .$^b$ & 60    & . . . & . . . \\
			COINS J2355+4950 & 0.238 & GPS   & G     & 1470  & 43.00 & 7     & 8     & 40.74   & 35    & . . . & . . . & 43.11 & $1.80^{+1.60}_{-0.90}$   & 60    & . . . & . . . \\
			B2 0026+34$^{\clubsuit}$ & 0.517 & GPS   & G     & 1310  & 43.65 & 7     & 333   & 43.06 & 36    & . . . & . . . & 44.26 & $1.43^{+0.20}_{-0.19}$  & 64    & 8.57 & -2.80 \\
			B3 0710+439 & 0.518 & GPS   & Q     & 1670  & 43.76 & 7     & 200   & 42.84 &  23    & 2032 & 43.66 & 44.52 & $1.59^{+0.07}_{-0.07}$  & 62    & . . .  & . . .\\
			PKS 0941-08 & 0.228 & GPS   & G     & 1100  & 42.84 & 7     & . . . & . . . &  . . . & . . . & . . . & 41.69 & $2.62^{+1.29}_{-1.03}$  & 43    & . . . & . . . \\
			SDSS J103507.04+562846.7 & 0.460  & GPS   & G     & 1270  & 43.53 & 7     & . . . &  . . . & . . . & 1829 & 43.50 & 43.33 & 1.75 & 60    & 9.37 & -3.38 \\
			4C 62.22$^{\clubsuit}$ & 0.429 & GPS   & G     & 1800  & 43.62  & 7     & 38    & 41.95 &  32    & 4375 & 43.81 & 44.47 & $1.24^{+0.17}_{-0.17}$  & 60    & 8.55 & -2.43 \\
			Mrk 0668 & 0.077 & GPS   & Q     & 2660  & 42.25 & 7     & . . . & . . . &  . . . & 830 & 41.49 & 43.95 & $2.21^{+0.19}_{-0.14}$  & 62    & 8.72 & -1.97 \\
			PKS 1718-649 & 0.014 & GPS   & G     & . . . & . . . & . . . & . . . &  . . . & . . . & . . . & . . . & 41.19 & $1.60^{+0.20}_{-0.20}$   & 62    & . . . & . . . \\
			COINS J1815+6127 & 0.601 & GPS   & Q     & . . . & . . . & . . . & . . . &  . . . & . . . & . . . & . . . & 44.04 & $1.70^{+0.33}_{-0.31}$   & 43    & . . . & . . . \\
			PKS 1934-63 & 0.183 & GPS   & G     & . . . & . . . & . . . & . . . &  . . . & . . . & . . . & . . . & 43.04 & $1.67^{+0.15}_{-0.16}$  & 62    & . . .  & . . . \\
			PKS 2127+04 & 0.990  & GPS   & G     & 2020  & 44.41 & 7     & 20    & 42.41   & 33    & 3867 & 44.57 & 44.45 & $1.98^{+0.50}_{-0.40}$  & 60    & . . . & . . . \\
			PKS 2254-367 & 0.006 & GPS   & G     & . . . & . . . & . . . & 530   & 39.32 & 37    & . . . & . . . & 40.90  & $1.89^{+0.10}_{-0.11}$  & 65    & . . . & . . . \\
			CGRaBs J1424+2256 & 3.620  & HFP   & Q     & . . . & . . . & . . . & . . . &  . . . & . . . & 285 & 44.90 & . . . & . . .  & . . . & . . . & . . . \\
			SDSS J130941.51+404757.2 & 2.907 & HFP   & Q     & 131   & 44.38 & 9     & . . . &  . . . & . . . & 39 & 43.85 & . . . & . . . & . . .     & 8.07 & 0.09 \\
			SDSS J151141.26+051809.2$^{\clubsuit}$ & 0.084 & HFP   & G     & 607   & 41.69 & 10    & 19$^{\ast}$  & 40.17  & 38    & 77 & 40.54 & 42.70  & $1.00^{+0.20}_{-0.20}$     & 62    & 7.88 & -2.23 \\
			CGRaBs J0111+3906 & 0.669 & HFP   & G     & 1270  & 43.87 & 11    & 229 & 43.24 & 19 & . . . & . . . & 43.83 & 1.75 & 66    & . . .  & . . . \\
			B2 0035+22 & 0.096 & CSO   & G     & . . . & . . . & . . . & 36 & 40.60 & 19 & . . . & . . . & 41.91 & 1.70  & 62   & . . . & . . . \\
			NGC 262 & 0.015 & CSO   & G  & 802 & 40.30  & 12    & 220   & 39.74 &  39    & . . . & . . . & . . . & . . . & . . .    & . . . & . . . \\
			NGC 3894 & 0.010  & CSO   & G     & 568   & 39.44  & 13    & 67    & 38.87 &  13    & 472 & 39.72 & 40.77 & $1.23^{+1.90}_{-0.65}$  & 67    & . . . & . . . \\
			SDSS J124733.31+672316.4 & 0.107 & CSO   & G     & . . . & . . . & . . . &  . . . & . . . & . . . & . . . & . . . & . . . & . . . & . . .     & 8.95 & -3.92 \\
			Mrk 231 & 0.042 & CSO   & G     & 270   & 40.73 & 14    & 162   & 40.51 &  14    & 243 & 40.41 & 42.48 & 1.47 & 68    & . . . & . . . \\
			PKS 1413+135 & 0.247 & CSO   & G     & 1000  & 42.87 & 15    & 547   & 42.61 &  40    & 1179 & 42.71 & 44.10  & $1.59^{+0.23}_{-0.22}$  & 42    & . . . & . . . \\
			COINS J1845+3541 & 0.763 & CSO   & G     & . . . & . . . & . . . & 173   & 43.12 &  41    & . . . & . . . & 43.70  & 1.70  & 62    & . . . & . . . \\
			COINS J1944+5448 & 0.263 & CSO   & G     & 885   & 42.87 & 16    & 14    & 41.07 &  35    & . . . & . . . & 42.87 & 1.70  & 62   & . . . & . . . \\
			COINS J2022+6136 & 0.227 & CSO   & G     & 2820  & 43.24 & 7     & 91    & 41.75  &  35    & . . . & . . . & 43.98 & $1.70^{+0.20}_{-0.10}$   & 63 & . . .  & . . .\\      
			\hline
		\end{tabular}
		\begin{minipage}{170mm}
			Columns: (1) source name; (2) redshift; (3) type; (4) ID: G - galaxy, Q - quasar; (5) - (6) the VLA core flux density (in mJy) and luminosity (in $\rm erg~s^{-1}$) at 5 GHz; (7) the references for (5); (8) - (9) the VLBI core flux density (in mJy) and luminosity (in $\rm erg~s^{-1}$) at 5 GHz; (10) the references for (8); (11) - (12) the FIRST 1.4 GHz flux density (in mJy), and corresponding luminosity at 5 GHz (assuming a spectral index of 0.5, in $\rm erg~s^{-1}$); (13) - (14) the X-ray luminosity (in $\rm erg~s^{-1}$) and the photon index at 2$-$10 keV; (15) the references for (13) and (14); (16) - (17): the black hole mass in solar mass, and the Eddington ratio $L_{\rm bol}/L_{\rm edd}$.\\
			$\clubsuit$: the source was used to derive the fundamental plane relation either with VLBI core luminosity applied or with FIRST data used except B2 0026+34.\\
			$\spadesuit$: the source was used to compute the composite SED.\\
			$\ast$: the radio flux density at 8.4 GHz, and the corresponding 5 GHz luminosity are derived assuming a spectral index of 0.\\
			$a$: $\Gamma$ = 1.7 is assumed to calculate the 2$-$10 keV X-ray luminosity.\\
			$b$: The assumed $\Gamma$ was used to estimate X-ray luminosity in \cite{teng09}.\\
			
			References: (1)\cite{1985AJ.....90..738P}; (2)\cite{1980AJ.....85..659S};
			(3)\cite{LM1997}             ; (4)\cite{1995Akujor}        ;
			(5)\cite{2009ApJ...694..992L}; (6)\cite{2017ApJ...851...87O};            
			(7)\cite{1998sta}           ;
			(8)\cite{1984ApJ...287...41W}; (9)\cite{2010MNRAS.408.1075O};
			(10)\cite{2005tinti}          ; (11)\cite{1990};
			(12)\cite{2006AJ....132..546G}; (13)\cite{Taylor1996}        ; (14)\cite{1999ApJ...516..127U};
			(15)\cite{1981AJ}            ; (16)\cite{2001MNRAS.321...37S};
			(17)\cite{2002fanti}         ; (18)\cite{2010MNRAS.402...87A};
			(19)http://astrogeo.org/vlbiimages/;
			(20)\cite{man2010};            (21)\cite{2006PASJ...58.1033S};
			(22)\cite{2002AJ....123.1258H};
			(23)\cite{2008ApJS..175..314D};      (24)\cite{2010ApJ...718.1345K};
			(25)\cite{1989fanti};                (26)\cite{2017MNRAS.466..952A};
			(27)\cite{2018RAA....18..108Y};      (28)\cite{2009de};
			(29)\cite{2010AJ....139.2612G};      (30)\cite{2003G};
			(31)\cite{1984ApJ...276..480J};      (32)\cite{2013MNRAS.433..147D};
			(33)\cite{1997sta};                  (34)\cite{sta1999};
			(35)\cite{2003ApJ...589..733P};      (36)\cite{2000ApJS..131...95F};
			(37)\cite{2015MNRAS.448..252T};      (38)\cite{2012ApJS..198....5A};
			(39)\cite{1983NGC262};               (40)\cite{perlman1996};
			(41)\cite{1995X};                    (42)This work;
			(43)\cite{2008ApJ...684..811S};      (44)\cite{2013ApJS..206....7M};
			(45)\cite{2018ApJS..235...32S};      (46)\cite{2013ApJ...777...27J};
			(47)\cite{2014MNRAS.437.3063K};      (48)\cite{2018MNRAS.476.5535T};
			(49)\cite{2009ApJ...705.1356K};      (50)\cite{2012ApJS..203...31M};
			(51)\cite{2018MNRAS.474.1342M};      (52)\cite{2010ApJ...714..589M};
			(53)\cite{2012MNRAS.424.1774H};
			(54)\cite{2011MNRAS.415.2910D};      (55)\cite{2015ApJS..220....5M};              
			(56)\cite{2010ApJ...713L..11Z};      (57)\cite{2013ApJ...773...15W};
			(58)\cite{2006MNRAS.366..339B};      (59)\cite{Salvati2008};                 
			(60)\cite{teng09};
			(61)\cite{2016ApJ...829...37B};      (62)\cite{2016ApJ...823...57S}; 
			(63)\cite{2019ApJ...871...71S}    
			(64)\cite{guainazzi06};
			(65)\cite{GM06};                     (66)\cite{2006MNRAS.367..928V};
			(67)\cite{2017ApJ...835..223S};      (68)\cite{2015Fer}
		\end{minipage}
	\end{center}
\end{table*}

\begin{table}
	\caption{Samples in our work and Fan $\&$ Bai (2016)}
	\begin{center}
		\begin{tabular}{lc}
			\hline \hline
			Sample & Source number\\
			\hline
			Our work & 91  \\
			Fan $\&$ Bai & 32  \\
			sources for FP in our work & 13\\
			sources for FP in Fan $\&$ Bai & 18\\
			common sources & 31$^{a}$  \\
			common sources for FP & 6  \\
			\hline	
		\end{tabular}
		\begin{minipage}{100mm}
			$a$: one source (0500+019) in Fan $\&$ Bai is excluded \\ in our work due to its blazar-type.
		\end{minipage}
	\end{center}	
\end{table}

\begin{table}
	\caption{Re-analysis of FP for Fan $\&$ Bai (2016) with corrected $L_{\rm X}$}
	\setlength{\tabcolsep}{0.05in}
	\begin{center}
		\begin{tabular}{lccccc}
			\hline \hline
			Sample & Number & b$^{a}$ & $\xi_{\rm RX}$$^{\ast}$ & $\xi_{\rm RM}$$^{\ast}$\\
			\hline
			All & 18 & $0.92\pm0.16$ & $0.66 \pm 0.08$ & $0.43 \pm 0.15$  \\
			All$^{b}$ & 18 &  & $0.81 \pm 0.07$ & $0.10 \pm 0.10$  \\
			Sources at $R_{\rm edd}$ $> 10^{-3}$  & 17$^{c}$ & $1.08\pm0.14$ & $0.92 \pm 0.05$   & $0.32 \pm 0.08$    \\
			Sources at $R_{\rm edd}$ $> 10^{-3}$ & 17$^{b,c}$&  & $0.95 \pm 0.06$   & $0.48 \pm 0.08$    \\
			\hline \hline
		\end{tabular}
		\begin{minipage}{90mm}
			$a$: using OLS bisector as did in our work.\\
			$\ast$: from same multiple linear regression of Bayesian
			approach (Kelly 2007) in our work and Fan $\&$ Bai (2016).\\
			$b$: the BH masses for 6 common sources have been replaced with ours.\\
			$c$: after excluding one source PKS 0941-08 with $L_{\rm bol}/L_{\rm edd}$ $< 10^{-3}$.
		\end{minipage}
	\end{center}	
\end{table}

{}

\bsp	
\label{lastpage}

\begin{thebibliography}{}

\bibitem[Akritas \& Siebert(1996)]{1996MNRAS.278..919A} Akritas, M.~G., \& Siebert, J.\ 1996, \mnras, 278, 919 \\

\bibitem[Akujor \& Garrington(1995)]{1995Akujor} Akujor, C.~E., \& Garrington, S.~T.\ 1995, \aaps, 112, 235 

\bibitem[An et al.(2010)]{2010MNRAS.402...87A} An, T., Hong, X.~Y., Hardcastle, M.~J., et al.\ 2010, \mnras, 402, 87 

\bibitem[An et al.(2017)]{2017MNRAS.466..952A} An, T., Lao, B.-Q., Zhao, W., et al.\ 2017, \mnras, 466, 952 

\bibitem[An et al.(2012)]{2012ApJS..198....5A} An, T., Wu, F., Yang, J., et al.\ 2012, \apjs, 198, 5 

\bibitem[Baker \& Hunstead(1995)]{1995baker} Baker, J.~C., \& Hunstead, R.~W.\ 1995, \apjl, 452, L95 

\bibitem[Barrows et al.(2016)]{2016ApJ...829...37B} Barrows, R.~S., Comerford, J.~M., Greene, J.~E., \& Pooley, D.\ 2016, \apj, 829, 37

\bibitem[Belsole et al.(2006)]{2006MNRAS.366..339B} Belsole, E., Worrall, D.~M., \& Hardcastle, M.~J.\ 2006, \mnras, 366, 339 \\

\bibitem[Bloom \& Marscher(1991)]{1991ApJ...366...16B} Bloom, S.~D., \& Marscher, A.~P.\ 1991, \apj, 366, 16 

\bibitem[Brandt, \& Alexander(2015)]{2015A&ARv..23....1B} Brandt, W.~N., \& Alexander, D.~M.\ 2015, \aapr, 23, 1

\bibitem[Chen(2018)]{2018ApJS..235...39C} Chen, L.\ 2018, \apjs, 235, 39


\bibitem[Coriat et al.(2011)]{2011MNRAS.414..677C} Coriat, M., Corbel, S., Prat, L., et al.\ 2011, \mnras, 414, 677 

\bibitem[Dallacasa et al.(2013)]{2013MNRAS.433..147D} Dallacasa, D., Orienti, M., Fanti, C., Fanti, R., \& Stanghellini, C.\ 2013, \mnras, 433, 147 

\bibitem[de Gasperin et al.(2011)]{2011MNRAS.415.2910D} de Gasperin, F., Merloni, A., Sell, P., et al.\ 2011, \mnras, 415, 2910

\bibitem[de Vries et al.(2009)]{2009de} de Vries, N., Snellen, I.~A.~G., Schilizzi, R.~T., Mack, K.-H., \& Kaiser, C.~R.\ 2009, \aap, 498, 641  

\bibitem[Dermer et al.(1992)]{1992A&A...256L..27D} Dermer, C.~D., Schlickeiser, R., \& Mastichiadis, A.\ 1992, \aap, 256, L27 

\bibitem[Dodson et al.(2008)]{2008ApJS..175..314D} Dodson, R., Fomalont, E.~B., Wiik, K., et al.\ 2008, \apjs, 175, 314 

\bibitem[Donato et al.(2001)]{2001A&A...375..739D} Donato, D., Ghisellini, G., Tagliaferri, G., et al.\ 2001, \aap, 375, 739

\bibitem[Dong et al.(2014)]{2014ApJ...787L..20D} Dong, A.-J., Wu, Q., \& Cao, X.-F.\ 2014, \apjl, 787, L20 

\bibitem[Fan \& Bai(2016)]{2016ApJ...818..185F} Fan, X.-L., \& Bai, J.-M.\ 2016, \apj, 818, 185

\bibitem[Fanaroff \& Riley(1974)]{1974FR} Fanaroff, B.~L., \& Riley, J.~M.\ 1974, \mnras, 167, 31P

\bibitem[Fanti et al.(2002)]{2002fanti} Fanti, C., Fanti, R., Dallacasa, D., et al.\ 2002, \aap, 396, 801 

\bibitem[Fanti et al.(1989)]{1989fanti} Fanti, C., Fanti, R., Parma, P., et al.\ 1989, \aap, 217, 44  

\bibitem[Fazio et al.(2004)]{2004ApJS..154...10F} Fazio, G.~G., Hora, J.~L., Allen, L.~E., et al.\ 2004, \apjs, 154, 10 

\bibitem[Feruglio et al.(2015)]{2015Fer} Feruglio, C., Fiore, F., Carniani, S., et al.\ 2015, \aap, 583, A99

\bibitem[Fomalont et al.(2000)]{2000ApJS..131...95F} Fomalont, E.~B., Frey, S., Paragi, Z., et al.\ 2000, \apjs, 131, 95 

\bibitem[Fruscione et al.(2006)]{2006SPIE.6270E..1VF} Fruscione, A., McDowell, J.~C., Allen, G.~E., et al.\ 2006, \procspie, 6270, 62701V 

\bibitem[Gallimore et al.(2006)]{2006AJ....132..546G} Gallimore, J.~F., Axon, D.~J., O'Dea, C.~P., Baum, S.~A., \& Pedlar, A.\ 2006, \aj, 132, 546 

\bibitem[Giroletti et al.(2003)]{2003G} Giroletti, M., Giovannini, G., Taylor, G.~B., et al.\ 2003, \aap, 399, 889 

\bibitem[Giroletti \& Polatidis(2009)]{2009AN....330..193G} Giroletti, M., \& Polatidis, A.\ 2009, Astronomische Nachrichten, 330, 193 
\\
\bibitem[Gonz{\'a}lez-Mart{\'{\i}}n et al.(2006)]{GM06} Gonz{\'a}lez-Mart{\'{\i}}n, O., Masegosa, J., M{\'a}rquez, I., Guerrero, M.~A., \& Dultzin-Hacyan, D.\ 2006, \aap, 460, 45 

\bibitem[Gu \& Chen(2010)]{2010AJ....139.2612G} Gu, M., \& Chen, Y.\ 2010, \aj, 139, 2612 


\bibitem[Guainazzi et al.(2006)]{guainazzi06} Guainazzi, M., Siemiginowska, A., Stanghellini, C., et al.\ 2006, \aap, 446, 87 


\bibitem[Hardcastle et al.(2012)]{2012MNRAS.424.1774H} Hardcastle, M.~J., Massaro, F., Harris, D.~E., et al.\ 2012, \mnras, 424, 1774 


\bibitem[Heinz \& Sunyaev(2003)]{2003MNRAS.343L..59H} Heinz, S., \& Sunyaev, R.~A.\ 2003, \mnras, 343, L59 

\bibitem[Helfand et al.(2015)]{2015ApJ...801...26H} Helfand, D.~J., White, R.~L., \& Becker, R.~H.\ 2015, \apj, 801, 26

\bibitem[Henley et al.(2010)]{2010ApJ...723..935H} Henley, D.~B., Shelton, R.~L., Kwak, K., Joung, M.~R., \& Mac Low, M.-M.\ 2010, \apj, 723, 935 


\bibitem[Hern{\'a}ndez-Garc{\'{\i}}a et al.(2015)]{2015A&A...579A..90H} Hern{\'a}ndez-Garc{\'{\i}}a, L., Masegosa, J., Gonz{\'a}lez-Mart{\'{\i}}n, O., \& M{\'a}rquez, I.\ 2015, \aap, 579, A90 

\bibitem[Hough et al.(2002)]{2002AJ....123.1258H} Hough, D.~H., Vermeulen, R.~C., Readhead, A.~C.~S., et al.\ 2002, \aj, 123, 1258 

\bibitem[\protect\citeauthoryear{Isobe, Feigelson, Akritas \& Babu}{1990}]{1990ApJ...364..104I} Isobe T., Feigelson E.~D., Akritas M.~G., Babu G.~J., 1990, ApJ, 364, 104 

\bibitem[Jia et al.(2013)]{2013ApJ...777...27J} Jia, J., Ptak, A., Heckman, T., \& Zakamska, N.~L.\ 2013, \apj, 777, 27 


\bibitem[Jones et al.(1984)]{1984ApJ...276..480J} Jones, D.~L., Wrobel, J.~M., \& Shaffer, D.~B.\ 1984, \apj, 276, 480 

\bibitem[Kalberla et al.(2005)]{2005A&A...440..775K} Kalberla, P.~M.~W., Burton, W.~B., Hartmann, D., et al.\ 2005, \aap, 440, 775

\bibitem[\protect\citeauthoryear{Kellermann, et al.}{1989}]{1989AJ.....98.1195K} Kellermann K.~I., Sramek R., Schmidt M., Shaffer D.~B., Green R., 1989, AJ, 98, 1195

\bibitem[Kelly(2007)]{2007ApJ...665.1489K} Kelly, B.~C.\ 2007, \apj, 665, 1489 


\bibitem[K{\"o}rding et al.(2006)]{2006A&A...456..439K} K{\"o}rding, E., Falcke, H., \& Corbel, S.\ 2006, \aap, 456, 439  

\bibitem[Kunert-Bajraszewska et al.(2010)]{2010ApJ...718.1345K} Kunert-Bajraszewska, M., Janiuk, A., Gawro{\'n}ski, M.~P., \& Siemiginowska, A.\ 2010, \apj, 718, 1345 

\bibitem[Kunert-Bajraszewska et al.(2014)]{2014MNRAS.437.3063K} Kunert-Bajraszewska, M., Labiano, A., Siemiginowska, A., \& Guainazzi, M.\ 2014, \mnras, 437, 3063

\bibitem[Kunert-Bajraszewska et al.(2009)]{2009ApJ...705.1356K} Kunert-Bajraszewska, M., Siemiginowska, A., Katarzy{\'n}ski, K., \& Janiuk, A.\ 2009, \apj, 705, 1356 

\bibitem[Laurent-Muehleisen et al.(1997)]{LM1997} Laurent-Muehleisen, S.~A., Kollgaard, R.~I., Ryan, P.~J., et al.\ 1997, \aaps, 122, 235 


\bibitem[Lawrence et al.(2012)]{2012yCat.2314....0L} Lawrence, A., Warren, S.~J., Almaini, O., et al.\ 2012, VizieR Online Data Catalog, 2314

\bibitem[Li(2019)]{2019MNRAS.490.3793L} Li, S.-L.\ 2019, \mnras, 490, 3793

\bibitem[Li \& Gu(2018)]{2018MNRAS.481L..45L} Li, S.-L., \& Gu, M.\ 2018, \mnras, 481, L45


\bibitem[Li et al.(2008)]{2008ApJ...688..826L} Li, Z.-Y., Wu, X.-B., \& Wang, R.\ 2008, \apj, 688, 826  


\bibitem[\protect\citeauthoryear{Liao \& Gu}{2020}]{paper1} Liao M., Gu M., 2020, MNRAS, 491, 92

\bibitem[Lin et al.(2009)]{2009ApJ...694..992L} Lin, Y.-T., Partridge, B., Pober, J.~C., et al.\ 2009, \apj, 694, 992 


\bibitem[Mantovani et al.(2010)]{man2010} Mantovani, F., Rossetti, A., Junor, W., Saikia, D.~J., \& Salter, C.~J.\ 2010, \aap, 518, A33 

\bibitem[Massaro et al.(2009)]{2009ApJ...692L.123M} Massaro, F., Chiaberge, M., Grandi, P., et al.\ 2009, \apjl, 692, L123 

\bibitem[Massaro et al.(2011)]{2011ApJS..197...24M} Massaro, F., Harris, D.~E., \& Cheung, C.~C.\ 2011, \apjs, 197, 24 

\bibitem[Massaro et al.(2015)]{2015ApJS..220....5M} Massaro, F., Harris, D.~E., Liuzzo, E., et al.\ 2015, \apjs, 220, 5

\bibitem[Massaro et al.(2010)]{2010ApJ...714..589M} Massaro, F., Harris, D.~E., Tremblay, G.~R., et al.\ 2010, \apj, 714, 589 

\bibitem[Massaro et al.(2013)]{2013ApJS..206....7M} Massaro, F., Harris, D.~E., Tremblay, G.~R., et al.\ 2013, \apjs, 206, 7

\bibitem[Massaro et al.(2012)]{2012ApJS..203...31M} Massaro, F., Tremblay, G.~R., Harris, D.~E., et al.\ 2012, \apjs, 203, 31 

\bibitem[Merloni et al.(2003)]{2003MNRAS.345.1057M} Merloni, A., Heinz, S., \& di Matteo, T.\ 2003, \mnras, 345, 1057 

\bibitem[Mezcua et al.(2018)]{2018MNRAS.474.1342M} Mezcua, M., Hlavacek-Larrondo, J., Lucey, J.~R., et al.\ 2018, \mnras, 474, 1342 

\bibitem[Migliori et al.(2012)]{2012ApJ...749..107M} Migliori, G., Siemiginowska, A., \& Celotti, A.\ 2012, \apj, 749, 107 


\bibitem[Migliori et al.(2014)]{2014ApJ...780..165M} Migliori, G., Siemiginowska, A., Kelly, B.~C., et al.\ 2014, \apj, 780, 165 

\bibitem[Morrissey et al.(2007)]{2007ApJS..173..682M} Morrissey, P., Conrow, T., Barlow, T.~A., et al.\ 2007, \apjs, 173, 682 

\bibitem[Murgia(2003)]{2003PASA...20...19M} Murgia, M.\ 2003, \pasa, 20, 19 

\bibitem[Neff \& de Bruyn(1983)]{1983NGC262} Neff, S.~G., \& de Bruyn, A.~G.\ 1983, \aap, 128, 318

\bibitem[\protect\citeauthoryear{Netzer}{2019}]{2019MNRAS.488.5185N} Netzer H., 2019, MNRAS, 488, 5185

\bibitem[O'Dea(1998)]{od98} O'Dea, C.~P.\ 1998, \pasp, 110, 493 

\bibitem[O'Dea et al.(1990)]{1990} O'Dea, C.~P., Baum, S.~A., Stanghellini, C., et al.\ 1990, \aaps, 84, 549 


\bibitem[O'Dea et al.(2017)]{2017ApJ...851...87O} O'Dea, C.~P., Worrall, D.~M., Tremblay, G.~R., et al.\ 2017, \apj, 851, 87 


\bibitem[Orienti et al.(2010)]{2010MNRAS.408.1075O} Orienti, M., Dallacasa, D., \& Stanghellini, C.\ 2010, \mnras, 408, 1075 


\bibitem[Ostorero et al.(2010)]{2010ApJ...715.1071O} Ostorero, L., Moderski, R., Stawarz, {\L}., et al.\ 2010, \apj, 715, 1071 

\bibitem[Owsianik et al.(1998)]{1998A&A...336L..37O} Owsianik, I., Conway, J.~E., \& Polatidis, A.~G.\ 1998, \aap, 336, L37 

\bibitem[Patnaik et al.(1992)]{1992MNRAS.259P...1P} Patnaik, A.~R., Browne, I.~W.~A., Walsh, D., Chaffee, F.~H., \& Foltz, C.~B.\ 1992, \mnras, 259, 1P 

\bibitem[Pearson et al.(1985)]{1985AJ.....90..738P} Pearson, T.~J., Perley, R.~A., \& Readhead, A.~C.~S.\ 1985, \aj, 90, 738 

\bibitem[Perlman et al.(1996)]{perlman1996} Perlman, E.~S., Carilli, C.~L., Stocke, J.~T., \& Conway, J.\ 1996, \aj, 111, 1839 

\bibitem[Polatidis \& Conway(2003)]{2003PASA...20...69P} Polatidis, A.~G., \& Conway, J.~E.\ 2003, \pasa, 20, 69 

\bibitem[Pollack et al.(2003)]{2003ApJ...589..733P} Pollack, L.~K., Taylor, G.~B., \& Zavala, R.~T.\ 2003, \apj, 589, 733 

\bibitem[Qiao \& Liu(2015)]{2015MNRAS.448.1099Q} Qiao, E., \& Liu, B.~F.\ 2015, \mnras, 448, 1099


\bibitem[Rieke et al.(2004)]{2004ApJS..154...25R} Rieke, G.~H., Young, E.~T., Engelbracht, C.~W., et al.\ 2004, \apjs, 154, 25 

\bibitem[Saikia et al.(2001)]{2001MNRAS.321...37S} Saikia, D.~J., Jeyakumar, S., Salter, C.~J., et al.\ 2001, \mnras, 321, 37 

\bibitem[Salvati et al.(2008)]{Salvati2008} Salvati, M., Risaliti, G., V{\'e}ron, P., \& Woltjer, L.\ 2008, \aap, 478, 121 

\bibitem[Schechter \& Moore(1993)]{Schechter1993} Schechter, P.~L., \& Moore, C.~B.\ 1993, \aj, 105, 1 

\bibitem[Shang et al.(2011)]{2011ApJS..196....2S} Shang, Z., Brotherton, M.~S., Wills, B.~J., et al.\ 2011, \apjs, 196, 2 

\bibitem[She et al.(2017)]{2017ApJ...835..223S} She, R., Ho, L.~C., \& Feng, H.\ 2017, \apj, 835, 223 

\bibitem[Shen et al.(2006)]{2006PASJ...58.1033S} Shen, Z.-Q., Shang, L.-L., Jiang, D.-R., Cai, H.-B., \& Chen, X.\ 2006, \pasj, 58, 1033 

\bibitem[Siemiginowska et al.(2008)]{2008ApJ...684..811S} Siemiginowska, A., LaMassa, S., Aldcroft, T.~L., Bechtold, J., \& Elvis, M.\ 2008, \apj, 684, 811

\bibitem[Siemiginowska et al.(2016)]{2016ApJ...823...57S} Siemiginowska, A., Sobolewska, M., Migliori, G., et al.\ 2016, \apj, 823, 57

\bibitem[Skrutskie et al.(2006)]{2006AJ....131.1163S} Skrutskie, M.~F., Cutri, R.~M., Stiening, R., et al.\ 2006, \aj, 131, 1163 

\bibitem[Snellen et al.(2003)]{2003PASA...20...38S} Snellen, I.~A.~G., Mack, K.-H., Schilizzi, R.~T., \& Tschager, W.\ 2003, \pasa, 20, 38  

\bibitem[Sobolewska et al.(2019)]{2019ApJ...871...71S} Sobolewska, M., Siemiginowska, A., Guainazzi, M., et al.\ 2019, \apj, 871, 71 


\bibitem[Spangler \& Cook(1980)]{1980AJ.....85..659S} Spangler, S.~R., \& Cook, D.~B.\ 1980, \aj, 85, 659 

\bibitem[Stanghellini et al.(1997)]{1997sta} Stanghellini, C., O'Dea, C.~P., Baum, S.~A., et al.\ 1997, \aap, 325, 943 

\bibitem[Stanghellini et al.(1998)]{1998sta} Stanghellini, C., O'Dea, C.~P., Dallacasa, D., et al.\ 1998, \aaps, 131, 303 \\

\bibitem[Stanghellini et al.(1999)]{sta1999} Stanghellini, C., O'Dea, C.~P., \& Murphy, D.~W.\ 1999, \aaps, 134, 309 

\bibitem[Stawarz et al.(2008)]{2008ApJ...680..911S} Stawarz, {\L}., Ostorero, L., Begelman, M.~C., et al.\ 2008, \apj, 680, 911 

\bibitem[Stuardi et al.(2018)]{2018ApJS..235...32S} Stuardi, C., Missaglia, V., Massaro, F., et al.\ 2018, \apjs, 235, 32 \\

\bibitem[Taylor et al.(1996)]{Taylor1996} Taylor, G.~B., Vermeulen, R.~C., Readhead, A.~C.~S., et al.\ 1996, \apjs, 107, 37 

\bibitem[Tengstrand et al.(2009)]{teng09} Tengstrand, O., Guainazzi, M., Siemiginowska, A., et al.\ 2009, \aap, 501, 89 

\bibitem[Tingay \& Edwards(2015)]{2015MNRAS.448..252T} Tingay, S.~J., \& Edwards, P.~G.\ 2015, \mnras, 448, 252 

\bibitem[Tinti et al.(2005)]{2005tinti} Tinti, S., Dallacasa, D., de Zotti, G., Celotti, A., \& Stanghellini, C.\ 2005, \aap, 432, 31 

\bibitem[Torresi et al.(2018)]{2018MNRAS.476.5535T} Torresi, E., Grandi, P., Capetti, A., Baldi, R.~D., \& Giovannini, G.\ 2018, \mnras, 476, 5535

\bibitem[Ulvestad et al.(1981)]{1981AJ} Ulvestad, J., Johnston, K., Perley, R., \& Fomalont, E.\ 1981, \aj, 86, 1010 

\bibitem[Ulvestad et al.(1999)]{1999ApJ...516..127U} Ulvestad, J.~S., Wrobel, J.~M., \& Carilli, C.~L.\ 1999, \apj, 516, 127 


\bibitem[Vink et al.(2006)]{2006MNRAS.367..928V} Vink, J., Snellen, I., Mack, K.-H., \& Schilizzi, R.\ 2006, \mnras, 367, 928 

\bibitem[Xu et al.(1995)]{1995X} Xu, W., Readhead, A.~C.~S., Pearson, T.~J., Polatidis, A.~G., \& Wilkinson, P.~N.\ 1995, \apjs, 99, 297 

\bibitem[Yuan et al.(2018)]{2018RAA....18..108Y} Yuan, Y., Gu, M.-F., \& Chen, Y.-J.\ 2018, Research in Astronomy and Astrophysics, 18, 108 

\bibitem[Wang et al.(2004)]{2004ApJ...607L.107W} Wang, J.-M., Watarai, K.-Y., \& Mineshige, S.\ 2004, \apjl, 607, L107 
\bibitem[Wilkes et al.(2013)]{2013ApJ...773...15W} Wilkes, B.~J., Kuraszkiewicz, J., Haas, M., et al.\ 2013, \apj, 773, 15

\bibitem[Worrall et al.(2004)]{2004MNRAS.347..632W} Worrall, D.~M., Hardcastle, M.~J., Pearson, T.~J., \& Readhead, A.~C.~S.\ 2004, \mnras, 347, 632  

\bibitem[Wright et al.(2010)]{2010AJ....140.1868W} Wright, E.~L., Eisenhardt, P.~R.~M., Mainzer, A.~K., et al.\ 2010, \aj, 140, 1868 

\bibitem[Wrobel \& Heeschen(1984)]{1984ApJ...287...41W} Wrobel, J.~M., \& Heeschen, D.~S.\ 1984, \apj, 287, 41 



\bibitem[Zhou \& Zhang(2010)]{2010ApJ...713L..11Z} Zhou, X.-L., \& Zhang, S.-N.\ 2010, \apjl, 713, L11 

\end{thebibliography}
\end{document}